\documentclass[prb,twocolumn,amsmath,amssymb,citeautoscript,showkeys,floatfix,superscriptaddress]{revtex4-2}

\usepackage{chemformula} % Formula subscripts using \ch{}
\usepackage[T1]{fontenc} % Use modern font encodings
\usepackage{siunitx}
\usepackage{amssymb}
\usepackage{amsmath}
\usepackage{graphicx}% http://ctan.org/pkg/graphicx
\usepackage{multirow,makecell}% http://ctan.org/pkg/multirow
\usepackage{booktabs}% http://ctan.org/pkg/booktabs
\usepackage{footnote}
\usepackage[toc,page]{appendix}

\usepackage[flushleft]{threeparttable}
\usepackage{comment}

\usepackage{hyperref}% add hypertext capabilities
\hypersetup{
    colorlinks,%
    citecolor=blue,%
    linkcolor=blue,%
    urlcolor=blue
}

\providecommand{\U}[1]{\protect\rule{.1in}{.1in}}

\date{\today}

\begin{document}

\title{Assessing the SCAN functional for deep defects and small polarons in wide-bandgap semiconductors and insulators}

\author{Darshana Wickramaratne}
\affiliation{Center for Computational Materials Science, US Naval Research Laboratory, Washington,
D.C. 20375, USA}
\author{John L. Lyons}
\affiliation{Center for Computational Materials Science, US Naval Research Laboratory, Washington,
D.C. 20375, USA}
\begin{abstract}
We find the recently developed strongly constrained and appropriately normed (SCAN) functional, now widely used in calculations of many materials, is not able to reliably describe the properties of deep defects and small polarons in a set of wide-bandgap semiconductors and insulators (ZnO, ZnSe, GaN, Ga$_2$O$_3$, and NaF). By comparing first-principles calculations using the SCAN functional against established experimental information and first-principles calculations using a hybrid functional, we find that the SCAN functional systematically underestimates the magnitude of the structural distortions at deep defects and tends to delocalize the charge density of these defect states.
\end{abstract}

\maketitle
\section{Introduction}
Density functional theory (DFT) calculations using supercells are the workhorse method for
investigating dopants and defects in semiconductors \cite{FreysoldtRMP2014}.  These studies play a central role in
defect identification and also in guiding the discovery of new materials and strategies to control their
conductivity.
Some examples of this include calculations of point defects in
III-nitrides for optoelectronics \cite{LyonsNCM2017}, transparent conducting oxides \cite{willis2021latest}
to defects in wide-bandgap semiconductors
for quantum information science \cite{ivady2018first,dreyer2018first}.

This success is due in part to the development and application of advanced functionals such as the Heyd, Scuseria, and Ernzerhof (HSE) \cite{HeydJCP2003,*HeydJCP2006} hybrid functional, which reduces the amount of self-interaction error that is inherent to semi-local functionals, and leads to an improved description of defects in semiconductors \cite{pacchioni2000theoretical,LyonsNCM2017}. Self-interaction errors lead to underestimated band gaps and a tendency to delocalize charge density \cite{mori2008localization}, precluding an accurate description of charge localization and local lattice distortions, which are pertinent in the context of defect calculations.  However, the large computational expense of hybrid functionals limits their widespread use for large system sizes. One approach to overcome this is to rely on functionals that improve upon the shortcomings of traditional functionals such as the generalized gradient approximation (GGA) and local density approximation (LDA) without incurring large computational costs.
A prominent example is the recently developed strongly constrained and appropriately normed (SCAN) functional \cite{sun2015strongly}, which has been shown to have less self-interaction error compared to the GGA and LDA functionals \cite{peng2017synergy}.

Naturally, this motivates the question on whether the SCAN functional can be reliably applied to investigate
the properties of defects in semiconductors. Indeed, several studies have begun to explore this question~\cite{sun2016accurate,rijal2021charged,rushchanskii2021ordering,rauch2021defect,wexler2021optimizing,han2022self,maciaszek2023application,ivanov2023electronic}.
For the calculation of defect transition levels, one approach has been optimizing the structure with the SCAN functional and then performing a single shot (i.e., no atomic relaxations) self-consistent calculation with the HSE functional \cite{wexler2021optimizing}, while other studies have relied on a forthright application of SCAN to determine the atomic geometry and total energies \cite{rauch2021defect,sun2016accurate,rijal2021charged,rushchanskii2021ordering}.
These approaches assume the local distortions due to the defect in a given charge
state are correctly captured with SCAN.  The SCAN functional has also been applied successfully to determine
intra-defect transition energies and optical lineshapes of defects \cite{maciaszek2023application,ivanov2023electronic}.
However, the success of this approach is predicated on the single particle states of the defect residing within the gap, combined with the fact that SCAN yields reliable lattice dynamics for a wide range
of materials \cite{ning2022reliable}.
The overall consensus from these studies has been in favor of applying
the SCAN functional as a reliable means to study defects in semiconductors.

While the SCAN functional leads to an encouraging improvement over GGA and LDA in the description of semiconductors, in this paper we scrutinize a set of critical issues that have not been considered when applying the SCAN functional to the investigation of deep-level defects and small polarons. Deep defects are characterized by a transition level that is located in the gap (at an energy that makes room-temperature ionization unlikely) and a wavefunction that is localized on the length scale of an atomic bond. A small polaron forms when an excess charge carrier in a semiconductor or insulator introduces a lattice distortion that is large enough to lead to charge localization often around a single atomic site. We focus on seven defects that allow us to highlight the importance of charge localization and symmetry-breaking structural distortions $-$ aspects that are germane to the study of deep defects and small polarons in all gapped materials. The defects we will investigate are: substitutional calcium acceptors (Ca$_{\rm Ga}$) in GaN, substitional carbon acceptors (C$_{\rm N}$) in GaN, the zinc vacancy  ($V_{\rm Zn}$) and oxygen vacancy ($V_{\rm O}$) in ZnO, the zinc vacancy ($V_{\rm Zn}$) in ZnSe, substitutional magnesium acceptors (Mg$_{\rm Ga}$) in $\beta$-Ga$_2$O$_3$ and the self-trapped hole (hole polaron) in NaF. Each of these candidate defects have been identified in experiment, which allows us to benchmark our SCAN and HSE calculations.

We limit our investigation to semiconductors with band gaps greater than $\sim$3 eV.
Focusing on defects in materials with wide band gaps reduces the ambiguity in assessing the properties of deep defect levels, despite the presence of self-interaction errors. This is because the single particle levels of the defect are more likely to reside in the gap, instead of being hybridized with the band states of the host material.  This enables an evaluation of the other properties (charge localization, structural distortions, and relaxation energies due to optical transitions) that we are interested in as part of this study.

We find that while the SCAN functional does identify deep transition levels in the gap compared to GGA, it fails to identify asymmetric structural distortions or describe charge localization. Based on these insights,
we recommend that defect and doping studies that are based on the popular SCAN functional must be interpreted with caution.

\section{Results and Discussion}
\begin{figure}[!t]
\includegraphics[width=8.5cm]{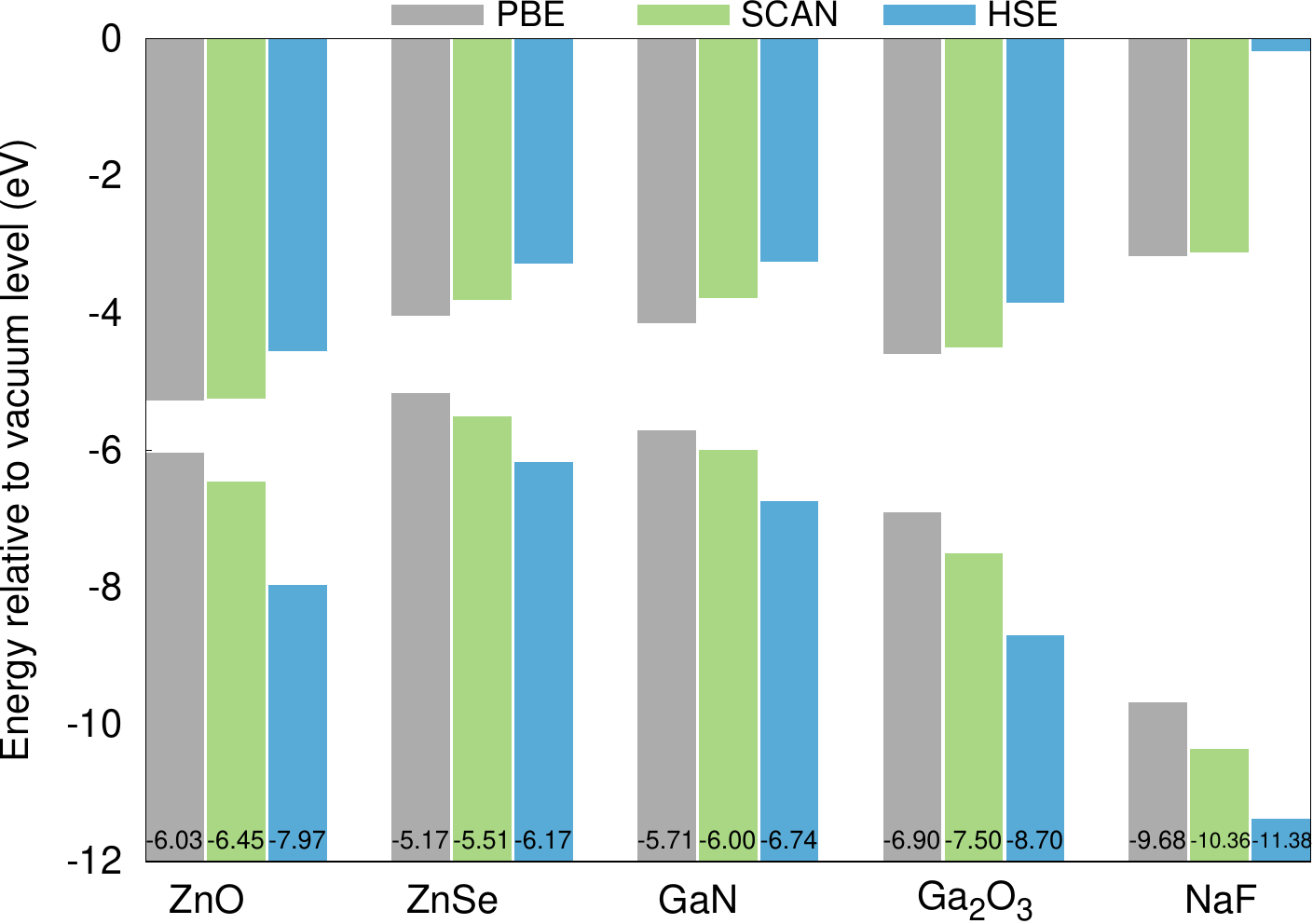}
\caption{
Absolute position of valence and conduction band edges with respect to the vacuum level using
PBE (grey), SCAN (green) and HSE (blue).
The valence band energies are denoted at the
bottom of each bar.}
\label{fig:align}%
\end{figure}

Table~\ref{table:bandgap} compares the lattice parameters and band gap of
GaN, ZnSe, ZnO, Ga$_2$O$_3$ and NaF obtained using our first-principles calculations (Appendix \ref{appendix:methods}), along
with experimental values.
\begin{table*}[]
\begin{center}
\caption{Lattice parameters and fundamental band gaps of ZnO, ZnSe, GaN, Ga$_2$O$_3$ and NaF obtained using PBE, SCAN and HSE.
Experimental results (Expt.) are from Ref.~\cite{madelung2012semiconductors} for ZnO, Ref.~\cite{passler1999temperature} for ZnSe, Ref.~\cite{maruska1969preparation,freitas2001structural} for GaN, Ref.~\cite{aahman1996reinvestigation,onuma2015valence} for Ga$_2$O$_3$
and Ref.~\cite{prencipe1995ab} for NaF.}
\begin{threeparttable}
\setlength{\tabcolsep}{6pt} % Default value: 6pt
\renewcommand{\arraystretch}{1.4} % Default value: 1
\begin{tabular}{lcccc|cccc}
  \toprule\toprule
  \multirow{2}{*}{\raisebox{-\heavyrulewidth}{Material}} & \multicolumn{4}{c}{Lattice parameters (\AA)} & \multicolumn{4}{c}{Band gap (eV)}\\
  \cmidrule(lr){2-9}
  & PBE & SCAN & HSE & Expt. & PBE & SCAN & HSE & Expt.\\
  \midrule
  ZnO & a=3.28,c=5.28 & a=3.23,c=5.21 & a=3.24,c=5.19  & a=3.25,c=5.21  & 0.76 & 1.21 & 3.42 & 3.43 \\
ZnSe & a=5.74 & a=5.63 & a=5.69 & a=5.67 & 1.14 & 1.70 & 2.90 & 2.82 \\
  GaN & a=3.22,c=5.22 & a=3.19,c=5.23 & a=3.19,c=5.17 & a=3.19,c=5.19 & 1.57 & 2.22 & 3.49 & 3.50  \\
  Ga$_2$O$_3$ & \multirowcell{2}{a=12.44,b=3.08,\\ c=5.92} & \multirowcell{2}{a=12.09,b=3.04,\\ c=5.82} & \multirowcell{2}{a=12.21,b=3.03,\\ c=5.80} & \multirowcell{2}{a=12.21,b=3.04,\\ c=5.80} & 2.25 & 2.94 & 4.86 & 4.48-4.90  \\ \\
  NaF & a=4.62 & a=4.47 & a=4.56 & a=4.57 & 6.51 & 7.25 & 11.20 & 11.50 \\
   \bottomrule\bottomrule
\end{tabular}
\label{table:bandgap}
\end{threeparttable}
\end{center}
\end{table*}
The SCAN functional yields structural parameters that are in good agreement with experiment, and leads to band gaps that are improved compared to the Perdew-Burke-Ernzerhof (PBE) functional, but still underestimated compared to experiment. For defect calculations, the absolute position of band edges is as important as the magnitude of the predicted band gap \cite{alkauskas2008defect}.  Figure \ref{fig:align} illustrates the absolute position of the band edges obtained
using surface slab calculations for each of the materials (Appendix \ref{appendix:align}).
Since the SCAN functional improves upon the description of band gaps and band edges compared to GGA, it does lead to charge-state
transition levels in the gap for the defects we investigate (whereas
some of these defects are predicted to be shallow with GGA).
When we compare our SCAN calculations with what is known experimentally about these defects, and with hybrid functional calculations, we note some key differences.

\begin{figure*}[!th]
\includegraphics[width=18.5cm]{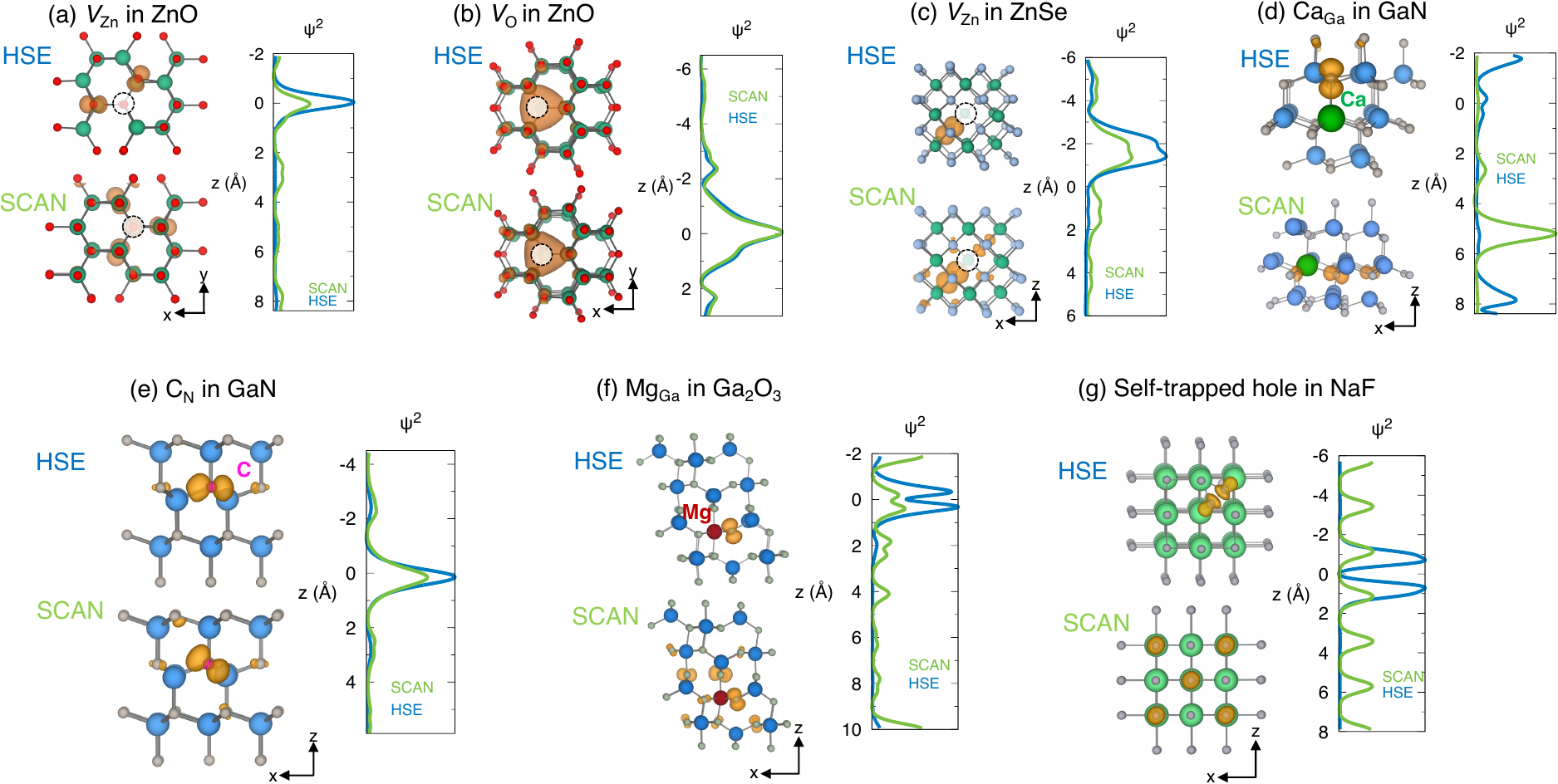}
\caption{Isosurfaces of the defect spin densities
obtained using HSE and SCAN for:
(a) $V_{\rm Zn}^{0}$ in ZnO,
(b) $V_{\rm O}^{+}$ in ZnO,
(c) $V_{\rm Zn}^{0}$ in ZnSe,
(d) Ca$_{\rm Ga}^{0}$ in GaN, (e) C$_{\rm N}^{0}$ in GaN, (f) Mg$_{\rm Ga}^{0}$ in Ga$_2$O$_3$, and
(g) the self-trapped hole in NaF.
The position of the vacancy in (a-c) is illustrated with a black dotted circle.
  A 1D profile of this density along the $z$-axis plotted with respect to the defect position
at $z$ = 0~\AA~is illustrated
to the right of the isosurface plots.}
\label{fig:density}%
\end{figure*}

The first case study is $V_{\rm Zn}$ in ZnO.  Prior HSE calculations \cite{frodason2017zn,Lyons17}
have shown $V_{\rm Zn}$ in ZnO exhibits a (2+/+) level at $E_v$+0.25 eV, a (+/0) level at $E_v$+0.91 eV,
a (0/$-$) level at $E_v$+1.45 eV and a ($-$/2$-$) level at $E_v$+1.92 eV (Appendix \ref{appendix:levels}), where $E_v$ is the ZnO valence-band maximum (VBM). With the SCAN functional, we find only the 0, $-$ and 2$-$ charge states in the gap, with the (0/$-$) level at $E_v$+0.25 eV and the ($-$/2$-$) level at $E_v$+0.51 eV. Furthermore, electron-paramagnetic resonance (EPR) experiments \cite{evans2008further} have suggested that each of the oxygen dangling bonds that are nearest neighbor to the $V_{\rm Zn}$ site can host a localized hole state.  This is corroborated by our HSE calculations [Fig.~\ref{fig:density}(a)], where we show holes localized on two of the oxygen dangling bonds in the neutral charge state of $V_{\rm Zn}$, consistent with the two electrons that can fill the Zn vacancy.  Localizing the density on each oxygen atom is
also accompanied by a reduction in the Zn-O bond lengths at the O atoms where charge is localized. In contrast, the SCAN functional delocalizes the charge density of the two empty states
across multiple O atoms [Fig.~\ref{fig:density}(a)] for the neutral charge state of $V_{\rm Zn}$.  This failure to describe charge localization also results in lower lattice distortions of the O atoms in the vicinity of the Zn vacancy site.

The second case study is the oxygen vacancy, $V_{\rm O}$, in ZnO.  Previous first-principles calculations have established
$V_{\rm O}$ is a deep donor in ZnO with a (2+/0) level in the gap \cite{oba2008defect,alkauskas2011band,Lyons17}.
The $q$=+ charge state of $V_{\rm O}$ is not stable
in equilibrium.  With HSE we find the (2+/0) level at $E_v$+1.96 eV, in agreement with prior HSE
calculations \cite{oba2008defect,Lyons17}.
With the SCAN functional we also find the (2+/0) level to be stable in the gap,
but at a significantly lower ionization energy of $E_v$+0.46 eV (Appendix \ref{appendix:levels}).
Calculations with HSE and SCAN find a breathing relaxation
of the Zn atoms that are nearest-neighbor to $V_{\rm O}$.
In the $q$=0 charge state, the Zn atoms move inwards by $\sim$9\%~with HSE and $\sim$12\%~with
the SCAN functional.  In the $q$=+ charge state, the Zn atoms move outwards by $\sim$4\%~with HSE and $\sim$2\%~with SCAN.
While the degree of charge localization is similar between HSE and SCAN for this defect (Figure \ref{fig:density}(b)),
we will show later that the
underestimated ionization energy and difference in relaxations between the two functionals quantitatively affects
comparisons between experimental signatures of this defect and first-principles calculations.

The third case study is the zinc vacancy, $V_{\rm Zn}$ in ZnSe, which acts as a deep acceptor analogous to the
case of $V_{\rm Zn}$ in ZnO.
The (0/$-$) level is at $E_v$+0.59 eV and the ($-$/2$-$) level is at $E_v$+1.07 eV with HSE.  We also find $V_{\rm Zn}$ can
trap an additional hole leading to a (+/0) level at $E_v$+0.21 eV.
With the SCAN functional, the acceptor levels are more shallow in comparison to HSE.  The (0/$-$) level
is at $E_v$+0.28 eV and the ($-$/2$-$) level is at $E_v$+0.64 eV (see Appendix \ref{appendix:levels}).
We find the SCAN functional fails to
stabilize the $+$ charge state in the gap.
In the negative charge state, where $V_{\rm Zn}$ hosts one hole,
EPR measurements have suggested the Se atom that localizes the hole distorts along the [111] direction
towards the vacancy site \cite{jeon1993optical}.
Our HSE calculations capture this structural distortion
and hole localization, as illustrated in Figure \ref{fig:density}(c).  The SCAN
calculations also lead to a lower symmetry structure. However, the magnitude of the distortions
are lower, and the charge density of the defect is delocalized to a greater degree beyond the Se atom
that is nearest-neighbor to the vacancy site, as compared to our HSE calculations (Figure \ref{fig:density}(c)).

The fourth case study is the Ca$_{\rm Ga}$ impurity in GaN.  It is a deep acceptor \cite{shen2017calcium} and has recently been identified as an efficient nonradiative recombination center in GaN \cite{young2016calcium,zhao2023trap}.
Both HSE and SCAN identify Ca$_{\rm Ga}$ as a deep acceptor: the (0/$-$) level is at $E_v$+0.73 eV with SCAN (Appendix \ref{appendix:levels}), and it is at $E_v$+1.06 eV using HSE (again, $E_v$ is the GaN VBM).  With HSE, we find Ca$_{\rm Ga}$ in the neutral charge state exhibits a large asymmetric distortion; the axial Ca-N bond length is 17\%~longer than the equilibrium Ga-N bond length, which localizes the hole on the axial N atom that is nearest neighbor to Ca$_{\rm Ga}$. Hole localization and large asymmetric distortions have been identified in the case of other group-II acceptors on the Ga site in GaN using HSE calculations \cite{lyons2013impact}.
With the SCAN functional, the nearest neighbor Ca-N bonds are symmetrically distorted in the neutral charge state, even if one starts from the HSE geometry with the asymmetric distortion.  Consequently, the hole state is delocalized across multiple nitrogen lattice sites for the calculations done with SCAN [Fig.~\ref{fig:density}(d)].

The fifth case study is C$_{\rm N}$ in GaN, which has been established as a deep acceptor \cite{kanegae2021photoionization, lyons2010carbon}.
HSE calculations find the (0/$-$) level at $E_v$+1.02 eV and with our SCAN calculations
the (0/$-$) level is more shallow at $E_v$+0.64 eV.
With HSE we find C$_{\rm N}$ in GaN can trap an additional hole leading to a (+/0) level at $E_v$+0.33 eV.  The SCAN functional
fails to stabilize
the positive charge state of C$_{\rm N}$.
The hole in the neutral charge state that arises from C$_{\rm N}$ is composed of C $2p$ orbitals, which we find in the gap in both HSE and SCAN calculations as illustrated in Figure~\ref{fig:density}(e). Note there is some delocalization of the charge density towards the second nearest neighbor N atoms predicted with the SCAN functional.  While both functionals show C$_{\rm N}$ acts as an atomic-like deep acceptor with the defect state derived from the impurity, the local lattice relaxation around the carbon impurity site differs between the two functionals.  With HSE, we find an asymmetric relaxation of the nearest neighbor C-Ga bonds in the neutral charge state, with one of the basal plane C-Ga bonds increasing by 6\%~relative to the equilibrium Ga-N bond length and the remaining three C-Ga bonds increasing by 1\%.
With our SCAN calculations, we find lower and symmetric distortions in the neutral charge state; the four nearest neighbor C-Ga bond lengths increase by 1.5\%. In the negative charge state the four C-Ga bonds decrease in length by 1\%~using HSE and SCAN.

The sixth case study is the Mg$_{\rm Ga}$ impurity, which is a deep acceptor in Ga$_2$O$_3$ \cite{bhandari2022optical}.
This is confirmed by both HSE and SCAN calculations.
The (0/$-$) level is at $E_v$+1.45 eV using HSE, which is close to the experimental value of 1.2 eV \cite{bhandari2022optical}.  With the SCAN functional, the (0/$-$) level is only at $E_v$+0.45 eV (Appendix \ref{appendix:levels}).  Mg$_{\rm Ga}^{0}$ is a polaronic acceptor since the hole
is localized primarily on an oxygen atom that is a nearest-neighbor to Mg$_{\rm Ga}$, and this is accompanied by a large lattice distortion of this Mg-O bond. This hole localization, which is captured in our HSE calculations [Fig.~\ref{fig:density}(f)], is in agreement with EPR experiments \cite{kananen2017electron} that identified evidence for hole trapping on the oxygen site due to Mg$_{\rm Ga}$. Hole localization at deep acceptors accompanied by large lattice distortions is a characteristic feature of several oxide semiconductors, and has proven to be crucial to understand the electrical and optical properties of this important class of materials \cite{varley2012role,lyons2022self}.  The SCAN calculations
lead to a lower lattice distortion in the vicinity of Mg$_{\rm Ga}$, and the hole in the neutral charge state is delocalized across multiple oxygen sites.

The final case study is the self-trapped hole in NaF, which is an archetypal example of a small
polaron. Experiments and theory \cite{williams1990self,miceli2018nonempirical} have shown that when holes are introduced into NaF, they self trap to form small hole polarons which are accompanied by the displacement of two near-neighbor F atoms along the $\langle$110$\rangle$ direction towards each other.  In our HSE calculations, we find the dimerization of two F atoms with a F-F bond length of 1.94~\AA,~with the hole localizing on both dimerized F atoms [Figure \ref{fig:density}(g)].  The SCAN functional fails to stabilize the small polaron in NaF.  It does not find the dimerization of the F atoms, and also delocalizes the wavefunction of the hole state across all of the F atoms.

In the seven case studies illustrated in Fig.~\ref{fig:density}, we find that though the SCAN functional identifies
deep thermodynamic transition levels (albeit with ionization energies that are underestimated compared to HSE), it systematically fails to describe charge localization of the defect states and fails to capture the large local structural distortions in the vicinity of the defect or polaron site. If we compare the thermodynamic transition levels obtained with calculations using SCAN and HSE we note that they are approximately
aligned on an absolute scale if we account for the differences in band edge energies
obtained with the two functionals (Appendix \ref{appendix:levels}).
While the combination of corrected band edge positions in combination with the calculation of thermodynamic transition levels with SCAN
may lead to an improved description of ionization energies, we note an accurate description of charge localization is essential to accurately predict a range of properties, which includes but is not limited to: electron-phonon coupling due to defects \cite{turiansky2021nonrad}, radiative recombination coefficients \cite{dreyer2020radiative} and hyperfine parameters \cite{ivady2018first}. The delocalized charge density associated with the defect states obtained with the SCAN functional, will lead to inaccurate predictions from such calculations.

The inability to identify charge localization and large structural distortions impacts the coupling between the defect
and the lattice.  If the occupation of the defect level is changed either electrically
or optically, or if there is a change in the configuration of a defect wavefunction within the same charge state (i.e., an intra-defect transition), the initial and final configurations often correspond to different geometries of the defect. These local distortions of the defect geometry manifest themselves experimentally and therefore serve as ``fingerprints'' that can be used to identify the microscopic origins of unknown defects. Coupling between the atomic and electronic structure of the defect is best understood using configuration coordinate diagrams and the Franck-Condon approximation as discussed in Ref.~\cite{alkauskas2016tutorial}. Briefly, in the context of an optical absorption measurement that involves a defect in a given charge state and photoionization to a band edge, within the Franck-Condon approximation
the change in geometry associated with the change in charge state of the defect leads to a peak absorption energy
that differs from the ionization energy of the defect.  This difference in energies is referred to as the relaxation energy. Analogous arguments apply in the context of photoluminescence that involves a defect state and a band edge.

\begin{figure}[!htb]
\includegraphics[width=8.5cm]{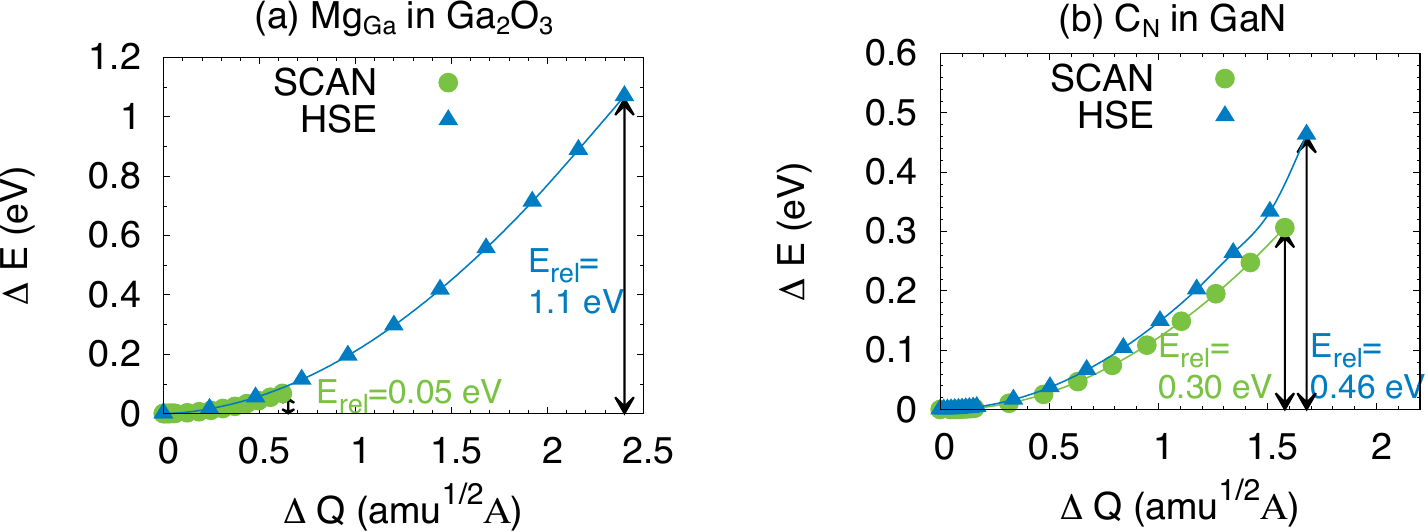}
\caption{Change in vibronic energy of (a) Mg$_{\rm Ga}$ in Ga$_2$O$_3$ and (b) C$_{\rm N}$ in GaN in their negative charge states.  The two sets of data points correspond to HSE (blue triangles) and SCAN (green circles) calculations illustrating the change in energy as the system transitions from the geometry of the negative charge state to that of the neutral charge state.}
\label{fig:ccd}%
\end{figure}

For example, for Mg$_{\rm Ga}$ in Ga$_2$O$_3$, experiments identified the relaxation energy associated with photoionization from the neutral to the negative charge state of Mg$_{\rm Ga}$ to be $\sim$1.3 eV, which is consistent with the relaxation energy of $\sim$1.1 eV we find in our HSE calculations of the configuration-coordinate diagram for this process illustrated in Fig.~\ref{fig:ccd}(a).
However, with the SCAN functional the change in atomic geometry between the two charge states is significantly lower than what we find with HSE, and the corresponding relaxation energy (0.05 eV) is more than an order of magnitude lower.

A similar situation occurs for $V_{\rm O}$ in ZnO, where optical absorption measurements in combination
with EPR have demonstrated evidence of photoionizing $V_{\rm O}^{0}$ and generating the EPR active charge state,
$V_{\rm O}^{+}$ \cite{wang2009oxygen}.  The metastable (+/0) level is 1.74 eV above the ZnO VBM with HSE,
and it is 0.12 eV above the ZnO VBM for our calculations with SCAN.
The optical process that stabilizes $V_{\rm O}^{+}$ occurs via optical
absorption with a peak at $\sim$ 2.6 eV, which is in good agreement with the peak absorption energy
predicted with HSE for the absorption process that converts $V_{\rm O}^{0}$ to $V_{\rm O}^{+}$ \cite{Lyons17}.
With the SCAN functional, the predicted peak absorption energy for this process is significantly lower; 1.79 eV.  This is due in part
to the underestimated ionization energy and underestimated lattice relaxation between the two charge states obtained
with the SCAN functional.

We also examined the photoluminescence process involving C$_{\rm N}$ in GaN where experiments \cite{ogino1980mechanism}
have shown a broad yellow luminescence band with a peak energy of $\sim$2.2 eV,
which is in good agreement with HSE calculations that predict a peak emission energy of 2.14 eV and a large relaxation energy of $\sim$0.5 eV \cite{lyons2010carbon}. Our SCAN calculations underestimate this relaxation energy, which we find to be $\sim$0.3 eV as illustrated in Fig.~\ref{fig:ccd}(b). This is due to an underestimation in the lattice relaxation between the neutral and negative charge state of C$_{\rm N}$.

We have shown, in the context of using the SCAN functional to analyze optical transitions due to these three defects, one would fail to describe the microscopic details such as the width of the optical spectra, even if the underestimated ionization energies were corrected in a post-processing step to match the HSE or experimental value.  We also considered how ionic screening due to changes in atomic configuration can affect the calculated relaxation energy \cite{falletta2020finite,gake2020finite,kumagai2023finite}. Using the approach outlined in Ref.\cite{falletta2020finite} for the case of C$_{\rm N}$ in GaN, we find the correction to be on the order of $\sim$0.1 eV with HSE and SCAN,  which is lower than the difference in relaxation energies obtained using both functionals.  Hence, we expect our conclusions about the tendency for the SCAN functional to underestimate relaxation energies will remain unchanged if these effects are accounted for the cases of the $V_{\rm O}$ in ZnO or Mg$_{\rm Ga}$ in Ga$_2$O$_3$, for which
the differences in relaxation energies between HSE and SCAN are larger.

The underestimated defect ionization energies obtained with the SCAN functional are due to the
underestimated band gaps and incorrect position of band edges.
Previous studies have corrected the underestimation of the band gap and the spurious delocalization of the charge density of deep defect states either by applying a Hubbard-$U$ correction to the states of the host material \cite{long2020evaluating}, a correction to the localized states of the defect \cite{lany2009polaronic} or applying modified semi-local functionals that enforce piecewise linearity of the total energy with respect to changes in occupation \cite{falletta2022polarons}.  It might be tempting to conclude that a forthright application of a Hubbard-$U$ correction provides a solution to the shortcomings with applying SCAN for defect calculations. A suitable test case is the $V_{\rm Zn}$ defect in ZnO, where calculations using SCAN underestimates the ionization energies and leads to delocalized charge densities.  This is due in part to the interaction of the filled Zn $d$ band with the O $2p$ states of the valence band, which leads to a VBM that is too high on an absolute scale. Applying $U$=5 eV to the Zn-$d$ states (consistent with GGA+$U$ calculations of ZnO \cite{janotti2007native}) increases the SCAN+$U$ ZnO band gap from 1.21 eV to 1.95 eV, lowers the VBM by $\sim$0.2 eV relative to the SCAN VBM, and increases the ionization energy of the $V_{\rm Zn}$ (0/$-$) and ($-$/2$-$) levels.  However, we find this SCAN+$U$ correction still leads to charge that is delocalized across multiple oxygen atoms, in contradiction with experiment and HSE calculations.

\section{Conclusions}
In conclusion, we examined the SCAN functional when applied to investigations of point defects
and self-trapped holes (small polarons) in wide-bandgap semiconductors and insulators, and compared our results with experiment and HSE calculations. For deep-level defects, the SCAN functional identifies thermodynamic transition levels in the gap.  However, for each of the deep defects that we investigated, SCAN delocalizes the charge density of the defect state, fails to identify asymmetric structural relaxations, and underestimates the change in geometry due to the change in charge state. For self-trapped holes, the SCAN functional fails to describe charge localization and the large lattice relaxation that accompanies the formation of these small polarons. We note that these shortcomings that we identified with applying the SCAN functional to defects also applies to the recently developed regularized SCAN functional, $r^{2}$SCAN (Appendix \ref{appendix:r2scan}).  While our investigation is limited to seven candidates, we anticipate our conclusions on the failure to describe charge localization, asymmetric structural distortions, ionization energies and relaxation energies of defects with the SCAN functional will broadly apply to the studies of deep defects in other semiconductors.  For example, hole localization by impurities has been identified in several nitride and oxide semiconductors, \cite{varley2012role} and asymmetric distortions due to impurities occur in a number of semiconductors.

\begin{acknowledgements}
We acknowledge insightful discussions with Anderson Janotti and Chris Van de Walle.
This work was supported by the Office of Naval Research through the Naval Research Laboratory's Basic Research Program, and computations were performed at the DoD Major Shared Resource Centers at AFRL and the Army ERDC.
\end{acknowledgements}

\appendix
\section{Computational Methods}
\label{appendix:methods}
Our calculations are performed using density functional theory (DFT) \cite{HK64,KohnPR1965}
with projector-augmented wave (PAW) potentials \cite{BlochlPRB1994} as implemented in the Vienna Ab initio Simulation Package ({\sc VASP}) \cite{KressePRB1996,kresse1994ab}.
We use the strongly constrained and appropriately normed (SCAN) meta-GGA functional \cite{sun2015strongly}
and the Heyd-Scuseria-Ernzerhof hybrid functional \cite{HeydJCP2003,*HeydJCP2006}.
We also verified our conclusions are unchanged by performing
calculations with the $r^{2}$SCAN functional \cite{furness2020accurate}.
The energy cutoff for the plane-wave basis set is 500 eV.  For the HSE calculations, the fraction of nonlocal
Hartree-Fock exchange is set to 0.31 for GaN, 0.375 for ZnO and ZnSe, 0.32 for Ga$_2$O$_3$, and 0.55 for NaF; this results in band gaps and lattice parameters that are in agreement with experiment for each of the materials.
We use the following {\sc VASP} 5.4 PBE PAWs - Ga PAWs with $d$ states in the valence, Na PAW with ($3s^{1}$)
in the valence, Mg with ($3s^{2}$) in the valence, Ca with ($3p^{6}$,$4s^{2}$) in the valence,
and the standard Zn, O, F, C and N PAWs.
Defect formation energies and thermodynamic transition levels are calculated using the standard supercell approach \cite{FreysoldtRMP2014}.
We use 96 atom supercells for GaN, 192 atom supercells for ZnO,
120 atom supercells for $\beta$-Ga$_2$O$_3$, and 216 atom supercells for NaF and ZnSe.
Previous calculations have shown that these supercell sizes yield converged
formation energies and thermodynamic transition levels for each of the defects and materials
that we investigate here \cite{Lyons17, bhandari2022optical, shen2017calcium}.
When comparing the SCAN versus HSE results, we use the equilibrium lattice parameters obtained from each functional
to construct the supercell, and then optimize all atomic coordinates.
Structure optimization and self-consistent calculations with supercells are performed using a single special $k$-point (1/4,1/4,1/4).
Spin polarization is taken into account in all of the calculations.
We include the non-spherical contributions related to the gradient of the density within the PAW spheres.
We verified that our conclusions remain unchanged if we use a $\Gamma$-centered (2$\times$2$\times$2) $k$-point grid
for the structural relaxations and total energy calculations.
The band edge energies were determined using symmetric surface-slab calculations with nonpolar surface terminations.  Slabs that are at least 15~\AA~in length are used to ensure they represent the bulk limit, and we use 15~\AA~of vacuum along the $c$-axis.

\section{Calculating band alignments}
\label{appendix:align}
\begin{figure}[!h]
\includegraphics[width=8.5cm]{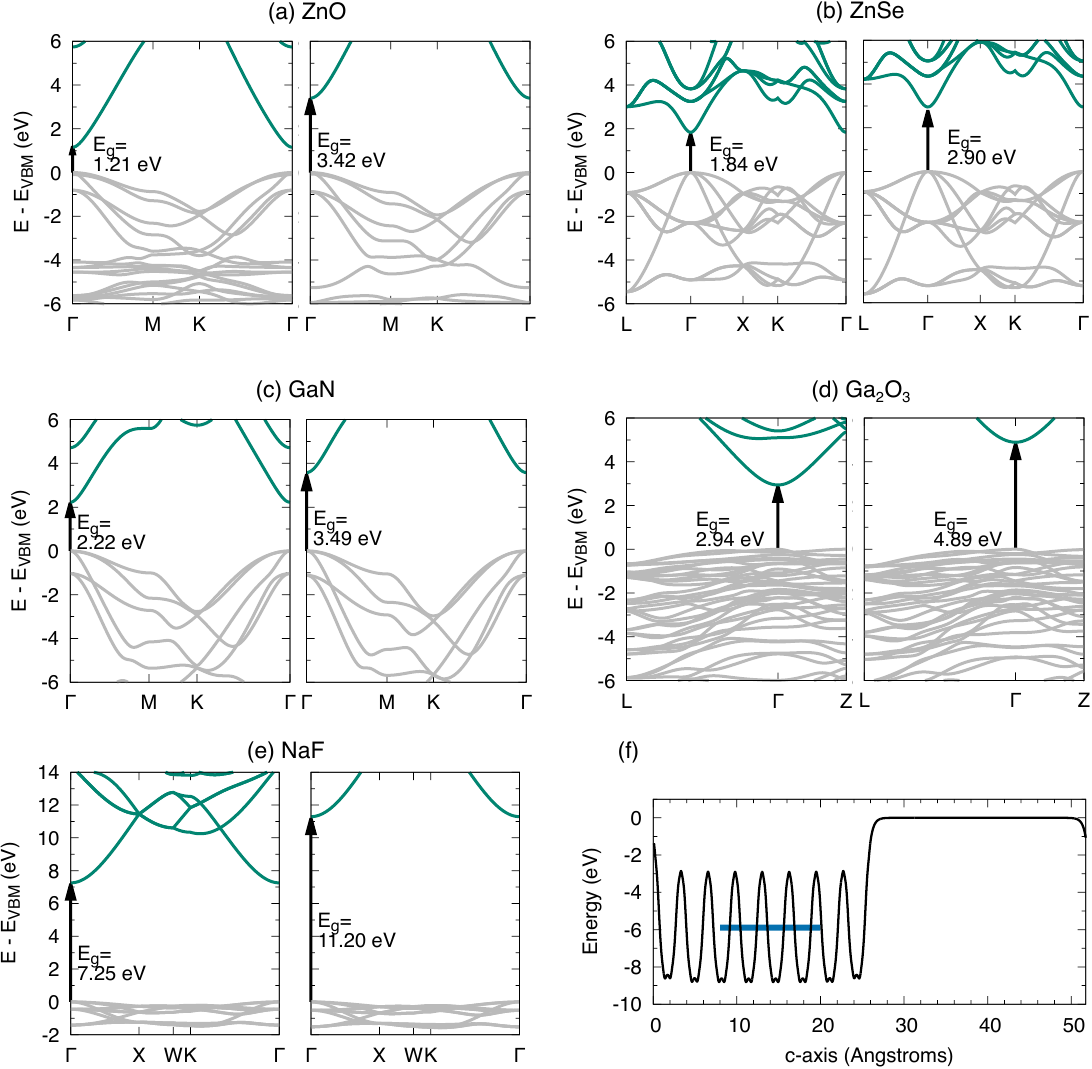}
\caption{HSE and SCAN band structures for (a) ZnO, (b) ZnSe, (c) GaN, (d) Ga$_2$O$_3$, and (e) NaF with the SCAN band
structure illustrated on the left and the HSE band structure on the right for each material.  (f) Planar
average of the electrostatic potential for ZnO obtained by averaging over planes that are parallel to the slab.
The blue line is the macroscopic average, obtained by averaging the planar-averaged potential over a bulk-like region
using the width of a single period.}
\label{fig:bands}%
\end{figure}

Bulk calculations of the band structure of each material using the SCAN and HSE functional
leads to the valence-band maximum (VBM) and conduction band minimum (CBM) energies
that are calculated with respect to the average electrostatic potential, $V_{\rm av}$.
The calculated band structures using the SCAN and HSE functionals for each of the materials is illustrated
in Figure \ref{fig:bands}.
Slab calculations using non-polar surfaces are used to determine the value of $V_{\rm av}$
with respect to the vacuum level by using a slab that is large enough so as to represent the bulk limit.
For GaN and ZnO we use (110) slabs, for Ga$_2$O$_3$ we use a (010) slab and for NaF we use a (100) slab.
In Figure \ref{fig:bands}(e) we illustrate the planar average of the electrostatic potential
for the ZnO slab calculation.  Performing a macroscopic average of this potential allows us to identify
the value of $V_{\rm av}$ with respect to the vacuum level (see blue horizontal line in Figure \ref{fig:bands}(e)).
This combination of bulk and slab calculations is then combined to obtain
the energy of the VBM and CBM with respect to the vacuum level.
Since we have the value of $V_{\rm av}$ referenced to the vacuum level from our slab calculations
this allows us to determine the natural band alignments where the VBM and CBM energies of each material
referenced to the vacuum level as illustrated in Figure 1 of the main text.

\section{Calculation of defect formation energies and thermodynamic transition levels}
\label{appendix:levels}
Obtaining the properties of the defects and small polarons reported in the main text
requires calculations of the formation energy.
For the different defects we investigate the formation energy of a defect in a charge state $q$
using the standard supercell approach as described in Ref.\cite{FreysoldtRMP2014}.  For the example, the formation energy of 
Ca$_{\rm Ga}$ in GaN is defined as:
\begin{equation}
\label{eqeform}
\begin{split}
E^f({\rm Ca}_{\rm Ga}^q)= E_{\rm tot}({\rm Ca}_{\rm Ga}^q) - E_{\rm tot}({\rm GaN}) + \mu_{\rm Ga} - \mu_{\rm Ca} \\ + q (E_F + \varepsilon_{v}) + \Delta^q
\end{split}
\end{equation}
in which $E_{\rm tot}$(Ca$_{\rm Ga}^q$) is the total energy of a supercell containing Ca$_{\rm Ga}^q$ in charge state $q$, and $E_{\rm tot}({\rm GaN})$ is the total energy of the same supercell without a defect. An electron added or removed from the supercell is exchanged with the Fermi level ($E_F$) of GaN which is referenced to the valence-band maximum (VBM) ($\varepsilon_v$).

The thermodynamic transition levels of the different defects corresponds to the Fermi energy where
the formation energy between two different charge states is equal.  For example the (0/$-$)
thermodynamic transition level of Ca$_{\rm Ga}$ in GaN with respect to the GaN VBM is defined as:
\begin{equation}
\label{eqetran}
(0/-) = E^f({\rm Ca}_{\rm Ga}^-;E_F=0) - E^f ({\rm Ca}_{\rm Ga}^0;E_F=0),
\end{equation}

\begin{figure}[!h]
\includegraphics[width=8.5cm]{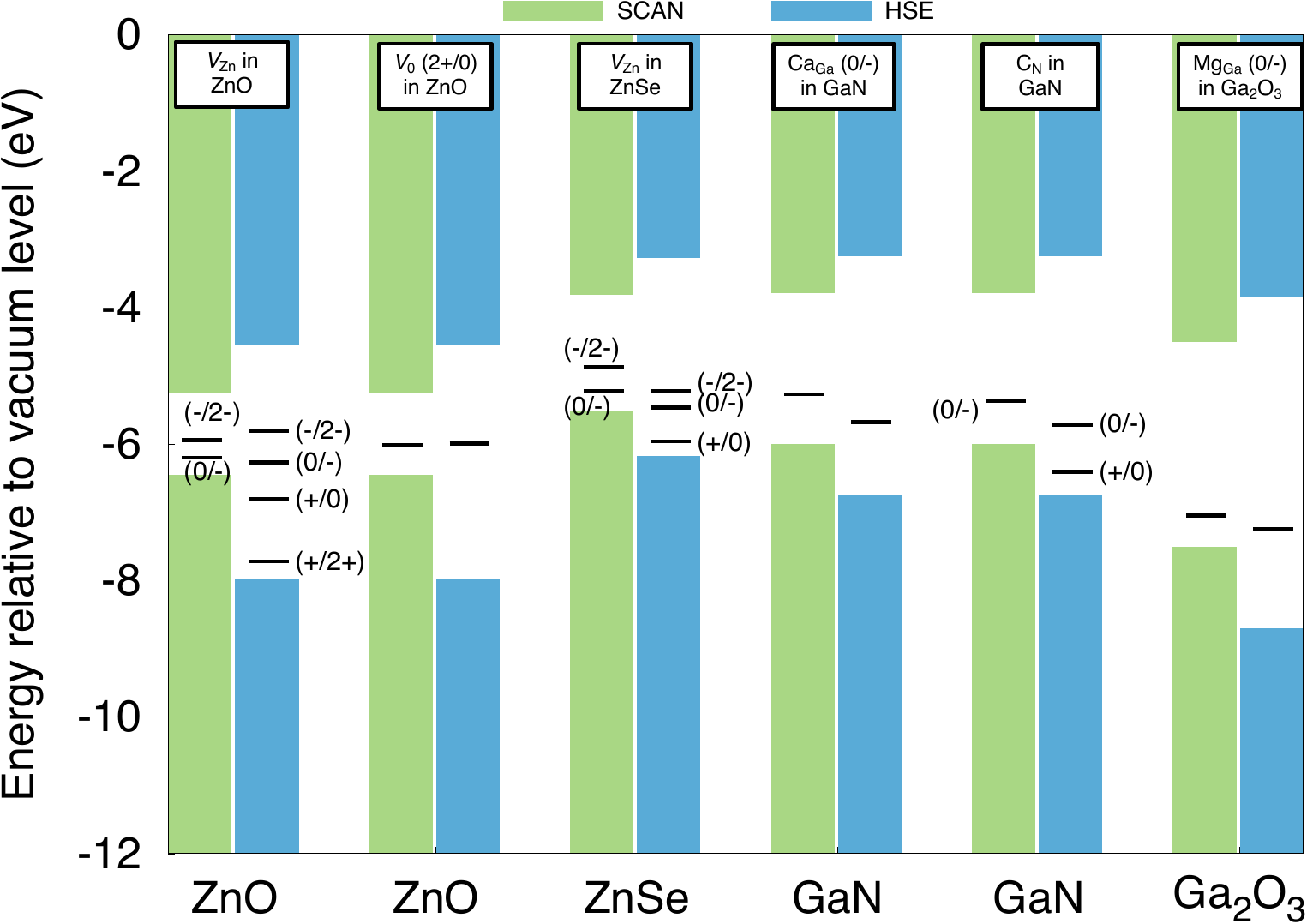}
\caption{Thermodynamic transition levels for $V_{\rm Zn}$ in ZnO, $V_{\rm O}$ in ZnO,
$V_{\rm Zn}$ in ZnSe,
Ca$_{\rm Ga}$ in GaN, C$_{\rm N}$ in GaN and Mg$_{\rm Ga}$ in Ga$_2$O$_3$ obtained
using the SCAN and HSE functional.}
\label{fig:levels}%
\end{figure}

The thermodynamic transition levels obtained with SCAN and HSE for the defects illustrated in
Figure 2 of the main text are illustrated in
Figure \ref{fig:levels}.

\section{Convergence with respect to $k$-point grid}
\label{appendix:kpt}
\begin{figure}[!h]
\includegraphics[width=8.5cm]{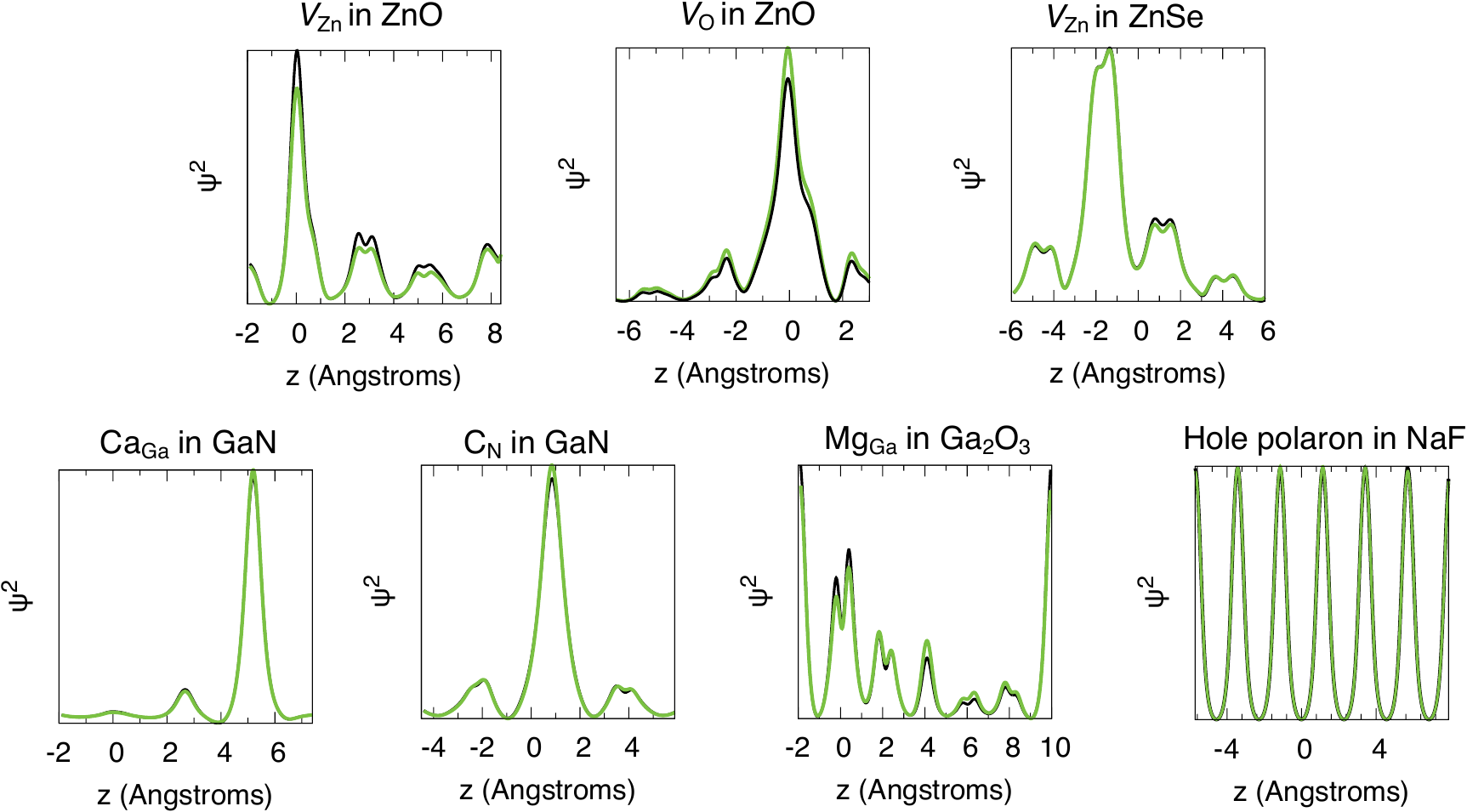}
\caption{
A 1D profile of this density along the $z$-axis plotted with respect to the defect position
at $z$ = 0~\AA~obtained using the SCAN functional for each of the defect configurations in Figure 2 of the main text.
Calculations using a single (1/4,1/4,1/4) $k$-point are illustrated in green and results
with a (2$\times$2$\times$2) $k$-point grid are illustrated in black.}
\label{fig:density_222}%
\end{figure}

It is important to consider whether the results we present in the main text
are affected by spurious effects due to the finite size supercells in our calculations.
First we consider the effect of $k$-point grid density that we use to optimize the supercell configurations
and determine the charge density of the defects states.  The results in the main text use a single
(1/4,1/4,1/4) $k$-point.
Using the SCAN functional, in Figure \ref{fig:density_222}, we compare the
planar average of the charge density of the defect states calculated with a (2$\times$2$\times$2) $k$-point grid
with the results in the main text.
The localization of the charge density using a (2$\times$2$\times$2) $k$-point grid and a single (1/4,1/4,1/4) $k$-point
are identical.

The atomic structure of the defects when optimizing with a single (1/4,1/4,1/4) $k$-point
versus the (2$\times$2$\times$2) $k$-point grid are also identical for this defect.
These tests demonstrate the results in the main text are converged with respect to
$k$-point grid.  We expect the same conclusions to apply to the HSE results reported in the main text.

\section{Impact of fraction of exact-exchange in hybrid functional calculations}
The fraction of exact-exchange that we employ in our HSE hybrid functional calculations is chosen
to match the experimental band gaps for each of the materials.  Using 25\%~exact exchange which was parametrized
for the HSE06 hybrid functional leads to underestimated band gaps for each of the materials that we
investigate.  This quantitatively affects the ionization energies that we calculate.
A pertinent question is whether using the default 25\%~mixing also affects the main focus of our study,
which is the ability to describe charge localization, asymmetric structural distortions and large structural
relaxation across each of the defects that we investigate here.

\begin{figure}[!h]
\includegraphics[width=8.5cm]{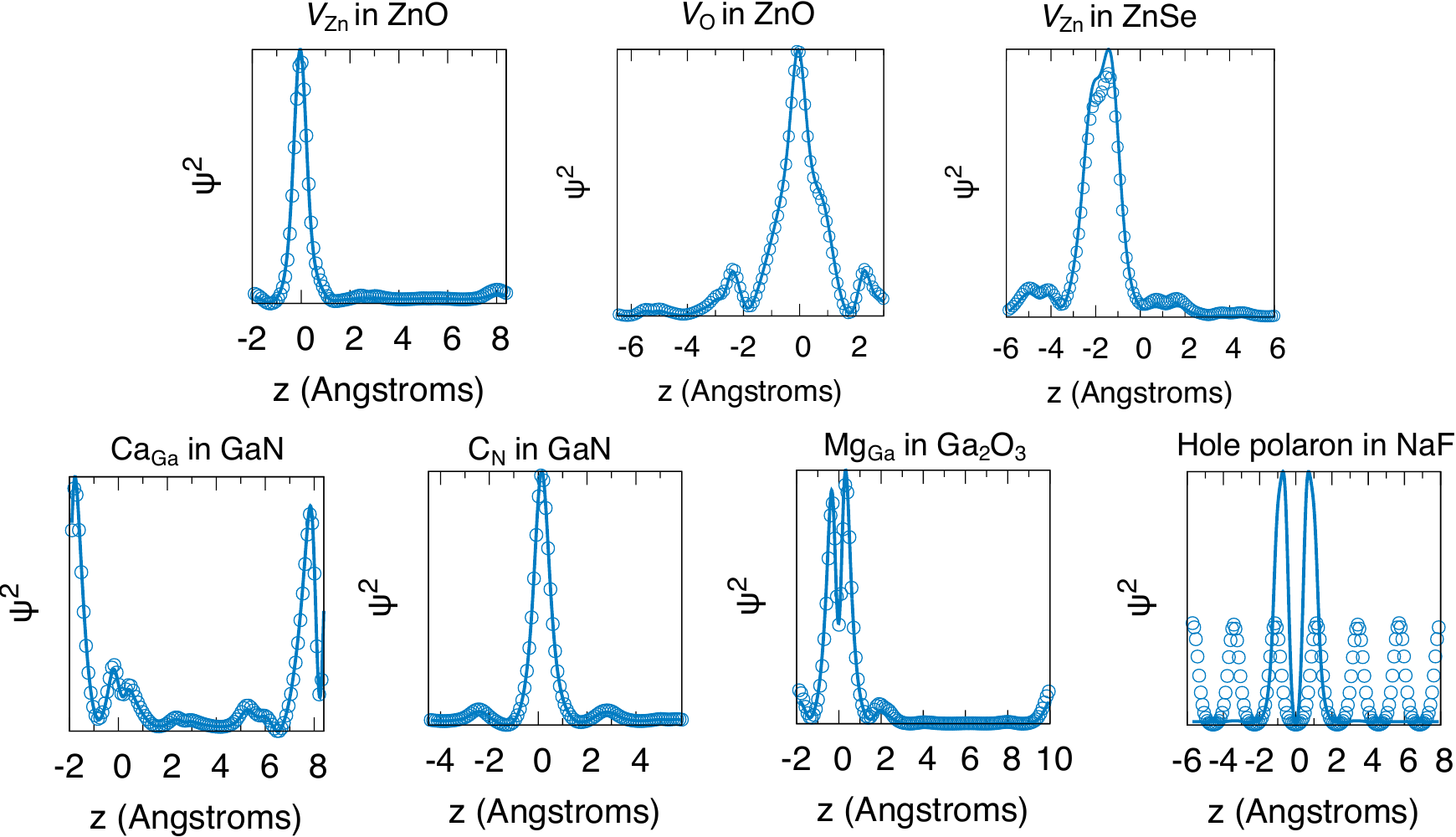}
\caption{
A 1D profile of this density along the $z$-axis plotted with respect to the defect position
at $z$ = 0~\AA~obtained
using the HSE hybrid functional.  Results from the main text that use an
optimally tuned fraction of exact exchange to match the experimental band gap are illustrated with blue solid lines
while results with the default HSE06 parameters (25\%~of exact exchange) are illustrated with the open circles.}
\label{fig:density25}%
\end{figure}

To verify this point, we optimized the atomic coordinates of all the defect/small polaron configurations
using the default HSE06 parameters (i.e. 25\%~of exact exchange).
In Figure \ref{fig:density25} we illustrate
the planar average of charge density of the defect states from each of these calculations.  For each
defect, we also compare against the results from the main text that rely on an optimally tuned
fraction of exact exchange chosen to match the material's experimental band gap.

The results in Figure \ref{fig:density25} clearly show that for the defects in ZnO, ZnSe, GaN and Ga$_2$O$_3$, the
degree of charge localization does not change for these defects if we use
the HSE06 parameters versus the mixing parameters that we list in Appendix A.
For NaF, we find the 25\%~of exact exchange with HSE06 is insufficient to lead to localization of the small polaron
in agreement with the conclusions of a prior study on the use of hybrid functionals to investigate small
polarons \cite{miceli2018nonempirical}.

\section{SCAN versus $r^{2}$SCAN}
\label{appendix:r2scan}
\begin{figure}[!bh]
\includegraphics[width=8.5cm]{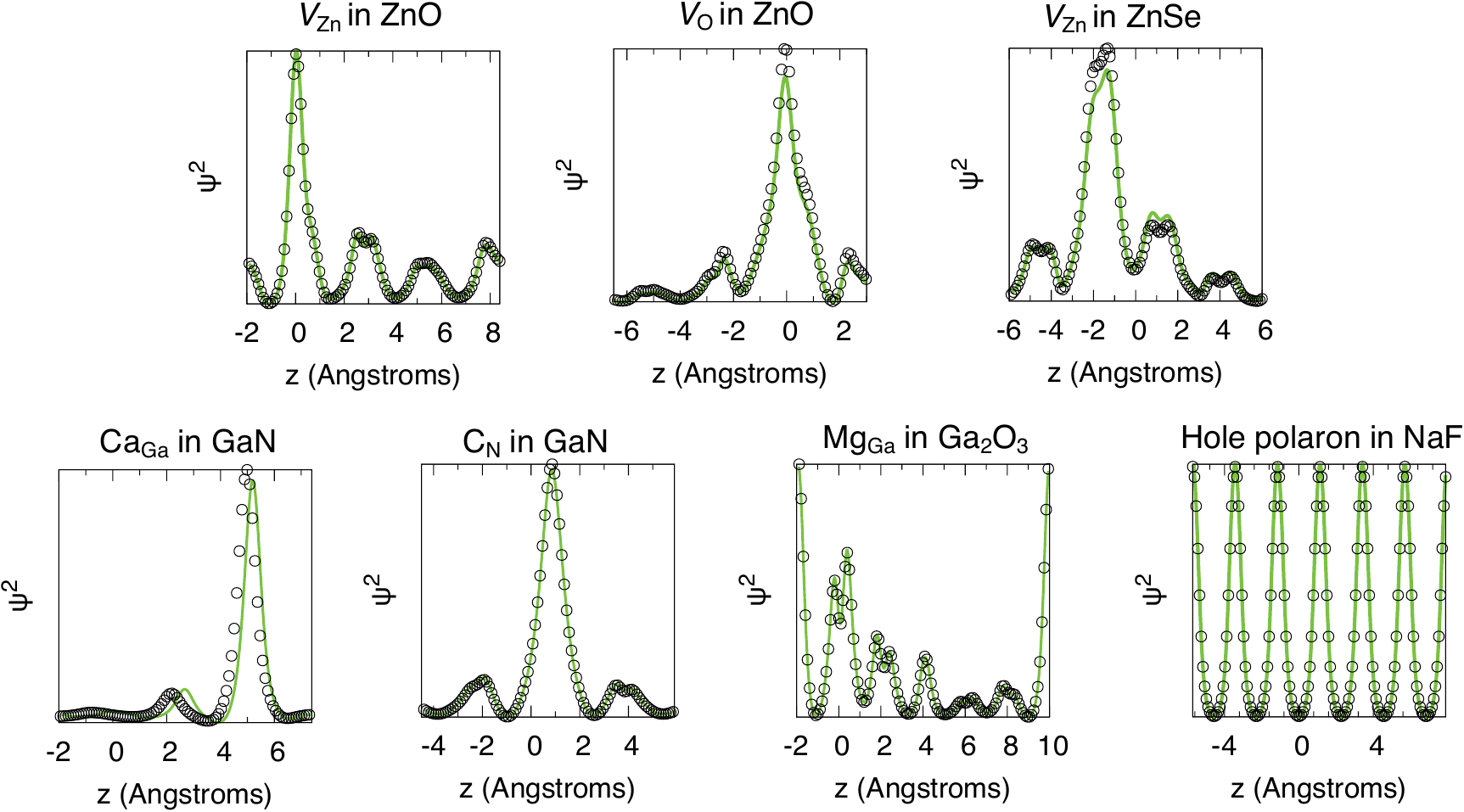}
\caption{
A 1D profile of this density along the $z$-axis plotted with respect to the defect position
at $z$ = 0~\AA~obtained
using the SCAN and $r^{2}$SCAN functional.  Results from the main text that use the SCAN functional are illustrated
with green solid lines while results with the $r^{2}$SCAN functional are illustrated with black open circles.}
\label{fig:r2scan}%
\end{figure}

One of the improvements to the SCAN functional has been the
introduction of the regularized SCAN functional ($r^{2}$SCAN) which improves upon
the numerical performance of the SCAN functional. We find SCAN and
$r^{2}$SCAN yield identical results as far as charge localization for the different defects
and small polaron configurations that we consider.  We optimized the atomic coordinates for
all of the defects/small polaron configurations using $r^{2}$SCAN and determine
the planar average of charge density of the defect states from each of these calculations.
In Figure \ref{fig:r2scan} we illustrate these results alongside the density obtained using SCAN.
Based on this it is clear our conclusions remain unchanged if the $r^{2}$SCAN functional is used.
We also verified the band gaps are nominally the same between SCAN and $r^{2}$SCAN in agreement with prior studies
that investigated the differences in electronic structure between the two approaches.

\newpage
%\bibliography{BIBLIO}

\begin{thebibliography}{64}%
\makeatletter
\providecommand \@ifxundefined [1]{%
 \@ifx{#1\undefined}
}%
\providecommand \@ifnum [1]{%
 \ifnum #1\expandafter \@firstoftwo
 \else \expandafter \@secondoftwo
 \fi
}%
\providecommand \@ifx [1]{%
 \ifx #1\expandafter \@firstoftwo
 \else \expandafter \@secondoftwo
 \fi
}%
\providecommand \natexlab [1]{#1}%
\providecommand \enquote  [1]{``#1''}%
\providecommand \bibnamefont  [1]{#1}%
\providecommand \bibfnamefont [1]{#1}%
\providecommand \citenamefont [1]{#1}%
\providecommand \href@noop [0]{\@secondoftwo}%
\providecommand \href [0]{\begingroup \@sanitize@url \@href}%
\providecommand \@href[1]{\@@startlink{#1}\@@href}%
\providecommand \@@href[1]{\endgroup#1\@@endlink}%
\providecommand \@sanitize@url [0]{\catcode `\\12\catcode `\$12\catcode
  `\&12\catcode `\#12\catcode `\^12\catcode `\_12\catcode `\%12\relax}%
\providecommand \@@startlink[1]{}%
\providecommand \@@endlink[0]{}%
\providecommand \url  [0]{\begingroup\@sanitize@url \@url }%
\providecommand \@url [1]{\endgroup\@href {#1}{\urlprefix }}%
\providecommand \urlprefix  [0]{URL }%
\providecommand \Eprint [0]{\href }%
\providecommand \doibase [0]{https://doi.org/}%
\providecommand \selectlanguage [0]{\@gobble}%
\providecommand \bibinfo  [0]{\@secondoftwo}%
\providecommand \bibfield  [0]{\@secondoftwo}%
\providecommand \translation [1]{[#1]}%
\providecommand \BibitemOpen [0]{}%
\providecommand \bibitemStop [0]{}%
\providecommand \bibitemNoStop [0]{.\EOS\space}%
\providecommand \EOS [0]{\spacefactor3000\relax}%
\providecommand \BibitemShut  [1]{\csname bibitem#1\endcsname}%
\let\auto@bib@innerbib\@empty
%</preamble>
\bibitem [{\citenamefont {Freysoldt}\ \emph {et~al.}(2014)\citenamefont
  {Freysoldt}, \citenamefont {Grabowski}, \citenamefont {Hickel}, \citenamefont
  {Neugebauer}, \citenamefont {Kresse}, \citenamefont {Janotti},\ and\
  \citenamefont {Van~de Walle}}]{FreysoldtRMP2014}%
  \BibitemOpen
  \bibfield  {author} {\bibinfo {author} {\bibfnamefont {C.}~\bibnamefont
  {Freysoldt}}, \bibinfo {author} {\bibfnamefont {B.}~\bibnamefont
  {Grabowski}}, \bibinfo {author} {\bibfnamefont {T.}~\bibnamefont {Hickel}},
  \bibinfo {author} {\bibfnamefont {J.}~\bibnamefont {Neugebauer}}, \bibinfo
  {author} {\bibfnamefont {G.}~\bibnamefont {Kresse}}, \bibinfo {author}
  {\bibfnamefont {A.}~\bibnamefont {Janotti}},\ and\ \bibinfo {author}
  {\bibfnamefont {C.~G.}\ \bibnamefont {Van~de Walle}},\ }\bibfield  {title}
  {\bibinfo {title} {First-principles calculations for point defects in
  solids},\ }\href {http://dx.doi.org/10.1103/RevModPhys.86.253} {\bibfield
  {journal} {\bibinfo  {journal} {Rev. Mod. Phys.}\ }\textbf {\bibinfo {volume}
  {86}},\ \bibinfo {pages} {253} (\bibinfo {year} {2014})}\BibitemShut
  {NoStop}%
\bibitem [{\citenamefont {Lyons}\ and\ \citenamefont {Van~de
  Walle}(2017)}]{LyonsNCM2017}%
  \BibitemOpen
  \bibfield  {author} {\bibinfo {author} {\bibfnamefont {J.~L.}\ \bibnamefont
  {Lyons}}\ and\ \bibinfo {author} {\bibfnamefont {C.~G.}\ \bibnamefont {Van~de
  Walle}},\ }\bibfield  {title} {\bibinfo {title} {Computationally predicted
  energies and properties of defects in {GaN}},\ }\href
  {https://doi.org/10.1038/s41524-017-0014-2} {\bibfield  {journal} {\bibinfo
  {journal} {npj Computational Mater.}\ }\textbf {\bibinfo {volume} {3}},\
  \bibinfo {pages} {1} (\bibinfo {year} {2017})}\BibitemShut {NoStop}%
\bibitem [{\citenamefont {Willis}\ and\ \citenamefont
  {Scanlon}(2021)}]{willis2021latest}%
  \BibitemOpen
  \bibfield  {author} {\bibinfo {author} {\bibfnamefont {J.}~\bibnamefont
  {Willis}}\ and\ \bibinfo {author} {\bibfnamefont {D.~O.}\ \bibnamefont
  {Scanlon}},\ }\bibfield  {title} {\bibinfo {title} {Latest directions in
  p-type transparent conductor design},\ }\href
  {https://doi.org/10.1039/D1TC02547C} {\bibfield  {journal} {\bibinfo
  {journal} {Journal of Materials Chemistry C}\ }\textbf {\bibinfo {volume}
  {9}},\ \bibinfo {pages} {11995} (\bibinfo {year} {2021})}\BibitemShut
  {NoStop}%
\bibitem [{\citenamefont {Iv{\'a}dy}\ \emph {et~al.}(2018)\citenamefont
  {Iv{\'a}dy}, \citenamefont {Abrikosov},\ and\ \citenamefont
  {Gali}}]{ivady2018first}%
  \BibitemOpen
  \bibfield  {author} {\bibinfo {author} {\bibfnamefont {V.}~\bibnamefont
  {Iv{\'a}dy}}, \bibinfo {author} {\bibfnamefont {I.~A.}\ \bibnamefont
  {Abrikosov}},\ and\ \bibinfo {author} {\bibfnamefont {A.}~\bibnamefont
  {Gali}},\ }\bibfield  {title} {\bibinfo {title} {First principles calculation
  of spin-related quantities for point defect qubit research},\ }\href
  {https://doi.org/10.1038/s41524-018-0132-5} {\bibfield  {journal} {\bibinfo
  {journal} {npj Computational Materials}\ }\textbf {\bibinfo {volume} {4}},\
  \bibinfo {pages} {76} (\bibinfo {year} {2018})}\BibitemShut {NoStop}%
\bibitem [{\citenamefont {Dreyer}\ \emph {et~al.}(2018)\citenamefont {Dreyer},
  \citenamefont {Alkauskas}, \citenamefont {Lyons}, \citenamefont {Janotti},\
  and\ \citenamefont {Van~de Walle}}]{dreyer2018first}%
  \BibitemOpen
  \bibfield  {author} {\bibinfo {author} {\bibfnamefont {C.~E.}\ \bibnamefont
  {Dreyer}}, \bibinfo {author} {\bibfnamefont {A.}~\bibnamefont {Alkauskas}},
  \bibinfo {author} {\bibfnamefont {J.~L.}\ \bibnamefont {Lyons}}, \bibinfo
  {author} {\bibfnamefont {A.}~\bibnamefont {Janotti}},\ and\ \bibinfo {author}
  {\bibfnamefont {C.~G.}\ \bibnamefont {Van~de Walle}},\ }\bibfield  {title}
  {\bibinfo {title} {First-principles calculations of point defects for quantum
  technologies},\ }\href {https://doi.org/10.1146/annurev-matsci-070317-124453}
  {\bibfield  {journal} {\bibinfo  {journal} {Annual Review of Materials
  Research}\ }\textbf {\bibinfo {volume} {48}},\ \bibinfo {pages} {1} (\bibinfo
  {year} {2018})}\BibitemShut {NoStop}%
\bibitem [{\citenamefont {Heyd}\ \emph {et~al.}(2003)\citenamefont {Heyd},
  \citenamefont {Scuseria},\ and\ \citenamefont {Ernzerhof}}]{HeydJCP2003}%
  \BibitemOpen
  \bibfield  {author} {\bibinfo {author} {\bibfnamefont {J.}~\bibnamefont
  {Heyd}}, \bibinfo {author} {\bibfnamefont {G.}~\bibnamefont {Scuseria}},\
  and\ \bibinfo {author} {\bibfnamefont {M.}~\bibnamefont {Ernzerhof}},\
  }\bibfield  {title} {\bibinfo {title} {{Hybrid functionals based on a
  screened Coulomb potential}},\ }\href {https://doi.org/{10.1063/1.1564060}}
  {\bibfield  {journal} {\bibinfo  {journal} {{J. Chem. Phys.}}\ }\textbf
  {\bibinfo {volume} {{118}}},\ \bibinfo {pages} {8207} (\bibinfo {year}
  {{2003}})}\BibitemShut {NoStop}%
\bibitem [{\citenamefont {Heyd}\ \emph {et~al.}(2006)\citenamefont {Heyd},
  \citenamefont {Scuseria},\ and\ \citenamefont {Ernzerhof}}]{HeydJCP2006}%
  \BibitemOpen
  \bibfield  {author} {\bibinfo {author} {\bibfnamefont {J.}~\bibnamefont
  {Heyd}}, \bibinfo {author} {\bibfnamefont {G.~E.}\ \bibnamefont {Scuseria}},\
  and\ \bibinfo {author} {\bibfnamefont {M.}~\bibnamefont {Ernzerhof}},\
  }\bibfield  {title} {\bibinfo {title} {Erratum: “hybrid functionals based
  on a screened coulomb potential” [j. chem. phys. 118, 8207 (2003)]},\
  }\href@noop {} {\bibfield  {journal} {\bibinfo  {journal} {J. Chem. Phys.}\
  }\textbf {\bibinfo {volume} {124}},\ \bibinfo {pages} {219906} (\bibinfo
  {year} {2006})}\BibitemShut {NoStop}%
\bibitem [{\citenamefont {Pacchioni}\ \emph {et~al.}(2000)\citenamefont
  {Pacchioni}, \citenamefont {Frigoli}, \citenamefont {Ricci},\ and\
  \citenamefont {Weil}}]{pacchioni2000theoretical}%
  \BibitemOpen
  \bibfield  {author} {\bibinfo {author} {\bibfnamefont {G.}~\bibnamefont
  {Pacchioni}}, \bibinfo {author} {\bibfnamefont {F.}~\bibnamefont {Frigoli}},
  \bibinfo {author} {\bibfnamefont {D.}~\bibnamefont {Ricci}},\ and\ \bibinfo
  {author} {\bibfnamefont {J.~A.}\ \bibnamefont {Weil}},\ }\bibfield  {title}
  {\bibinfo {title} {Theoretical description of hole localization in a quartz
  {Al} center: The importance of exact electron exchange},\ }\href@noop {}
  {\bibfield  {journal} {\bibinfo  {journal} {Phys. Rev. B}\ }\textbf {\bibinfo
  {volume} {63}},\ \bibinfo {pages} {054102} (\bibinfo {year}
  {2000})}\BibitemShut {NoStop}%
\bibitem [{\citenamefont {Mori-S{\'a}nchez}\ \emph {et~al.}(2008)\citenamefont
  {Mori-S{\'a}nchez}, \citenamefont {Cohen},\ and\ \citenamefont
  {Yang}}]{mori2008localization}%
  \BibitemOpen
  \bibfield  {author} {\bibinfo {author} {\bibfnamefont {P.}~\bibnamefont
  {Mori-S{\'a}nchez}}, \bibinfo {author} {\bibfnamefont {A.~J.}\ \bibnamefont
  {Cohen}},\ and\ \bibinfo {author} {\bibfnamefont {W.}~\bibnamefont {Yang}},\
  }\bibfield  {title} {\bibinfo {title} {Localization and delocalization errors
  in density functional theory and implications for band-gap prediction},\
  }\href {https://doi.org/10.1103/PhysRevLett.100.146401} {\bibfield  {journal}
  {\bibinfo  {journal} {Phys. Rev. Lett.}\ }\textbf {\bibinfo {volume} {100}},\
  \bibinfo {pages} {146401} (\bibinfo {year} {2008})}\BibitemShut {NoStop}%
\bibitem [{\citenamefont {Sun}\ \emph {et~al.}(2015)\citenamefont {Sun},
  \citenamefont {Ruzsinszky},\ and\ \citenamefont {Perdew}}]{sun2015strongly}%
  \BibitemOpen
  \bibfield  {author} {\bibinfo {author} {\bibfnamefont {J.}~\bibnamefont
  {Sun}}, \bibinfo {author} {\bibfnamefont {A.}~\bibnamefont {Ruzsinszky}},\
  and\ \bibinfo {author} {\bibfnamefont {J.~P.}\ \bibnamefont {Perdew}},\
  }\bibfield  {title} {\bibinfo {title} {Strongly constrained and appropriately
  normed semilocal density functional},\ }\href
  {https://doi.org/10.1103/PhysRevLett.115.036402} {\bibfield  {journal}
  {\bibinfo  {journal} {Phys. Rev. Lett.}\ }\textbf {\bibinfo {volume} {115}},\
  \bibinfo {pages} {036402} (\bibinfo {year} {2015})}\BibitemShut {NoStop}%
\bibitem [{\citenamefont {Peng}\ and\ \citenamefont
  {Perdew}(2017)}]{peng2017synergy}%
  \BibitemOpen
  \bibfield  {author} {\bibinfo {author} {\bibfnamefont {H.}~\bibnamefont
  {Peng}}\ and\ \bibinfo {author} {\bibfnamefont {J.~P.}\ \bibnamefont
  {Perdew}},\ }\bibfield  {title} {\bibinfo {title} {Synergy of van der {Waals}
  and self-interaction corrections in transition metal monoxides},\ }\href
  {https://doi.org/10.1103/PhysRevB.96.100101} {\bibfield  {journal} {\bibinfo
  {journal} {Phys. Rev. B}\ }\textbf {\bibinfo {volume} {96}},\ \bibinfo
  {pages} {100101} (\bibinfo {year} {2017})}\BibitemShut {NoStop}%
\bibitem [{\citenamefont {Sun}\ \emph {et~al.}(2016)\citenamefont {Sun},
  \citenamefont {Remsing}, \citenamefont {Zhang}, \citenamefont {Sun},
  \citenamefont {Ruzsinszky}, \citenamefont {Peng}, \citenamefont {Yang},
  \citenamefont {Paul}, \citenamefont {Waghmare}, \citenamefont {Wu} \emph
  {et~al.}}]{sun2016accurate}%
  \BibitemOpen
  \bibfield  {author} {\bibinfo {author} {\bibfnamefont {J.}~\bibnamefont
  {Sun}}, \bibinfo {author} {\bibfnamefont {R.~C.}\ \bibnamefont {Remsing}},
  \bibinfo {author} {\bibfnamefont {Y.}~\bibnamefont {Zhang}}, \bibinfo
  {author} {\bibfnamefont {Z.}~\bibnamefont {Sun}}, \bibinfo {author}
  {\bibfnamefont {A.}~\bibnamefont {Ruzsinszky}}, \bibinfo {author}
  {\bibfnamefont {H.}~\bibnamefont {Peng}}, \bibinfo {author} {\bibfnamefont
  {Z.}~\bibnamefont {Yang}}, \bibinfo {author} {\bibfnamefont {A.}~\bibnamefont
  {Paul}}, \bibinfo {author} {\bibfnamefont {U.}~\bibnamefont {Waghmare}},
  \bibinfo {author} {\bibfnamefont {X.}~\bibnamefont {Wu}}, \emph {et~al.},\
  }\bibfield  {title} {\bibinfo {title} {Accurate first-principles structures
  and energies of diversely bonded systems from an efficient density
  functional},\ }\href {https://doi.org/10.1038/nchem.2535} {\bibfield
  {journal} {\bibinfo  {journal} {Nat. Chem.}\ }\textbf {\bibinfo {volume}
  {8}},\ \bibinfo {pages} {831} (\bibinfo {year} {2016})}\BibitemShut {NoStop}%
\bibitem [{\citenamefont {Rijal}\ \emph {et~al.}(2021)\citenamefont {Rijal},
  \citenamefont {Tan}, \citenamefont {Freysoldt},\ and\ \citenamefont
  {Hennig}}]{rijal2021charged}%
  \BibitemOpen
  \bibfield  {author} {\bibinfo {author} {\bibfnamefont {B.}~\bibnamefont
  {Rijal}}, \bibinfo {author} {\bibfnamefont {A.~M.~Z.}\ \bibnamefont {Tan}},
  \bibinfo {author} {\bibfnamefont {C.}~\bibnamefont {Freysoldt}},\ and\
  \bibinfo {author} {\bibfnamefont {R.~G.}\ \bibnamefont {Hennig}},\ }\bibfield
   {title} {\bibinfo {title} {Charged vacancy defects in monolayer
  phosphorene},\ }\href {https://doi.org/10.1103/PhysRevMaterials.5.124004}
  {\bibfield  {journal} {\bibinfo  {journal} {Phys. Rev. Mater.}\ }\textbf
  {\bibinfo {volume} {5}},\ \bibinfo {pages} {124004} (\bibinfo {year}
  {2021})}\BibitemShut {NoStop}%
\bibitem [{\citenamefont {Rushchanskii}\ \emph {et~al.}(2021)\citenamefont
  {Rushchanskii}, \citenamefont {Bl{\"u}gel},\ and\ \citenamefont
  {Le{\v{z}}ai{\'c}}}]{rushchanskii2021ordering}%
  \BibitemOpen
  \bibfield  {author} {\bibinfo {author} {\bibfnamefont {K.~Z.}\ \bibnamefont
  {Rushchanskii}}, \bibinfo {author} {\bibfnamefont {S.}~\bibnamefont
  {Bl{\"u}gel}},\ and\ \bibinfo {author} {\bibfnamefont {M.}~\bibnamefont
  {Le{\v{z}}ai{\'c}}},\ }\bibfield  {title} {\bibinfo {title} {Ordering of
  oxygen vacancies and related ferroelectric properties in
  {HfO}$_{2-\delta}$},\ }\href {https://doi.org/10.1103/PhysRevLett.127.087602}
  {\bibfield  {journal} {\bibinfo  {journal} {Phys. Rev. Lett.}\ }\textbf
  {\bibinfo {volume} {127}},\ \bibinfo {pages} {087602} (\bibinfo {year}
  {2021})}\BibitemShut {NoStop}%
\bibitem [{\citenamefont {Rauch}\ \emph {et~al.}(2021)\citenamefont {Rauch},
  \citenamefont {Munoz}, \citenamefont {Marques},\ and\ \citenamefont
  {Botti}}]{rauch2021defect}%
  \BibitemOpen
  \bibfield  {author} {\bibinfo {author} {\bibfnamefont {T.}~\bibnamefont
  {Rauch}}, \bibinfo {author} {\bibfnamefont {F.}~\bibnamefont {Munoz}},
  \bibinfo {author} {\bibfnamefont {M.~A.}\ \bibnamefont {Marques}},\ and\
  \bibinfo {author} {\bibfnamefont {S.}~\bibnamefont {Botti}},\ }\bibfield
  {title} {\bibinfo {title} {Defect levels from {SCAN} and {mBJ meta-GGA}
  exchange-correlation potentials},\ }\href
  {https://doi.org/10.1103/PhysRevB.104.064105} {\bibfield  {journal} {\bibinfo
   {journal} {Phys. Rev. B}\ }\textbf {\bibinfo {volume} {104}},\ \bibinfo
  {pages} {064105} (\bibinfo {year} {2021})}\BibitemShut {NoStop}%
\bibitem [{\citenamefont {Wexler}\ \emph {et~al.}(2021)\citenamefont {Wexler},
  \citenamefont {Gautam},\ and\ \citenamefont {Carter}}]{wexler2021optimizing}%
  \BibitemOpen
  \bibfield  {author} {\bibinfo {author} {\bibfnamefont {R.~B.}\ \bibnamefont
  {Wexler}}, \bibinfo {author} {\bibfnamefont {G.~S.}\ \bibnamefont {Gautam}},\
  and\ \bibinfo {author} {\bibfnamefont {E.~A.}\ \bibnamefont {Carter}},\
  }\bibfield  {title} {\bibinfo {title} {Optimizing kesterite solar cells from
  {Cu$_2$ZnSnS$_4$} to {Cu$_2$CdGe(S, Se)$_4$}},\ }\href
  {https://doi.org/10.1039/D0TA11603C} {\bibfield  {journal} {\bibinfo
  {journal} {J. Mater. Chem. A}\ }\textbf {\bibinfo {volume} {9}},\ \bibinfo
  {pages} {9882} (\bibinfo {year} {2021})}\BibitemShut {NoStop}%
\bibitem [{\citenamefont {Han}\ \emph {et~al.}(2022)\citenamefont {Han},
  \citenamefont {Rudel}, \citenamefont {Schnick},\ and\ \citenamefont
  {Ebert}}]{han2022self}%
  \BibitemOpen
  \bibfield  {author} {\bibinfo {author} {\bibfnamefont {D.}~\bibnamefont
  {Han}}, \bibinfo {author} {\bibfnamefont {S.~S.}\ \bibnamefont {Rudel}},
  \bibinfo {author} {\bibfnamefont {W.}~\bibnamefont {Schnick}},\ and\ \bibinfo
  {author} {\bibfnamefont {H.}~\bibnamefont {Ebert}},\ }\bibfield  {title}
  {\bibinfo {title} {Self-doping behavior and cation disorder in {MgSnN$_2$}},\
  }\href {https://doi.org/10.1103/PhysRevB.105.125202} {\bibfield  {journal}
  {\bibinfo  {journal} {Phys. Rev. B}\ }\textbf {\bibinfo {volume} {105}},\
  \bibinfo {pages} {125202} (\bibinfo {year} {2022})}\BibitemShut {NoStop}%
\bibitem [{\citenamefont {Maciaszek}\ \emph {et~al.}(2023)\citenamefont
  {Maciaszek}, \citenamefont {{\v{Z}}alandauskas}, \citenamefont {Silkinis},
  \citenamefont {Alkauskas},\ and\ \citenamefont
  {Razinkovas}}]{maciaszek2023application}%
  \BibitemOpen
  \bibfield  {author} {\bibinfo {author} {\bibfnamefont {M.}~\bibnamefont
  {Maciaszek}}, \bibinfo {author} {\bibfnamefont {V.}~\bibnamefont
  {{\v{Z}}alandauskas}}, \bibinfo {author} {\bibfnamefont {R.}~\bibnamefont
  {Silkinis}}, \bibinfo {author} {\bibfnamefont {A.}~\bibnamefont
  {Alkauskas}},\ and\ \bibinfo {author} {\bibfnamefont {L.}~\bibnamefont
  {Razinkovas}},\ }\bibfield  {title} {\bibinfo {title} {The application of the
  {SCAN} density functional to color centers in diamond},\ }\href
  {https://doi.org/10.1063/5.0154319} {\bibfield  {journal} {\bibinfo
  {journal} {J. Chem. Phys.}\ }\textbf {\bibinfo {volume} {159}},\ \bibinfo
  {pages} {084708} (\bibinfo {year} {2023})}\BibitemShut {NoStop}%
\bibitem [{\citenamefont {Ivanov}\ \emph {et~al.}(2023)\citenamefont {Ivanov},
  \citenamefont {Schmerwitz}, \citenamefont {Levi},\ and\ \citenamefont
  {Jónsson}}]{ivanov2023electronic}%
  \BibitemOpen
  \bibfield  {author} {\bibinfo {author} {\bibfnamefont {A.~V.}\ \bibnamefont
  {Ivanov}}, \bibinfo {author} {\bibfnamefont {Y.~L.~A.}\ \bibnamefont
  {Schmerwitz}}, \bibinfo {author} {\bibfnamefont {G.}~\bibnamefont {Levi}},\
  and\ \bibinfo {author} {\bibfnamefont {H.}~\bibnamefont {Jónsson}},\
  }\bibfield  {title} {\bibinfo {title} {{Electronic excitations of the charged
  nitrogen-vacancy center in diamond obtained using time-independent
  variational density functional calculations}},\ }\href
  {https://doi.org/10.21468/SciPostPhys.15.1.009} {\bibfield  {journal}
  {\bibinfo  {journal} {SciPost Phys.}\ }\textbf {\bibinfo {volume} {15}},\
  \bibinfo {pages} {009} (\bibinfo {year} {2023})}\BibitemShut {NoStop}%
\bibitem [{\citenamefont {Ning}\ \emph {et~al.}(2022)\citenamefont {Ning},
  \citenamefont {Furness},\ and\ \citenamefont {Sun}}]{ning2022reliable}%
  \BibitemOpen
  \bibfield  {author} {\bibinfo {author} {\bibfnamefont {J.}~\bibnamefont
  {Ning}}, \bibinfo {author} {\bibfnamefont {J.~W.}\ \bibnamefont {Furness}},\
  and\ \bibinfo {author} {\bibfnamefont {J.}~\bibnamefont {Sun}},\ }\bibfield
  {title} {\bibinfo {title} {Reliable lattice dynamics from an efficient
  density functional approximation},\ }\href
  {https://doi.org/10.1021/acs.chemmater.1c03222} {\bibfield  {journal}
  {\bibinfo  {journal} {Chem. Mater.}\ }\textbf {\bibinfo {volume} {34}},\
  \bibinfo {pages} {2562} (\bibinfo {year} {2022})}\BibitemShut {NoStop}%
\bibitem [{\citenamefont {Madelung}(2012)}]{madelung2012semiconductors}%
  \BibitemOpen
  \bibfield  {author} {\bibinfo {author} {\bibfnamefont {O.}~\bibnamefont
  {Madelung}},\ }\href@noop {} {\emph {\bibinfo {title} {Semiconductors—basic
  data}}}\ (\bibinfo  {publisher} {Springer Science \& Business Media},\
  \bibinfo {year} {2012})\BibitemShut {NoStop}%
\bibitem [{\citenamefont {P{\"a}ssler}\ \emph {et~al.}(1999)\citenamefont
  {P{\"a}ssler}, \citenamefont {Griebl}, \citenamefont {Riepl}, \citenamefont
  {Lautner}, \citenamefont {Bauer}, \citenamefont {Preis}, \citenamefont
  {Gebhardt}, \citenamefont {Buda}, \citenamefont {As}, \citenamefont
  {Schikora} \emph {et~al.}}]{passler1999temperature}%
  \BibitemOpen
  \bibfield  {author} {\bibinfo {author} {\bibfnamefont {R.}~\bibnamefont
  {P{\"a}ssler}}, \bibinfo {author} {\bibfnamefont {E.}~\bibnamefont {Griebl}},
  \bibinfo {author} {\bibfnamefont {H.}~\bibnamefont {Riepl}}, \bibinfo
  {author} {\bibfnamefont {G.}~\bibnamefont {Lautner}}, \bibinfo {author}
  {\bibfnamefont {S.}~\bibnamefont {Bauer}}, \bibinfo {author} {\bibfnamefont
  {H.}~\bibnamefont {Preis}}, \bibinfo {author} {\bibfnamefont
  {W.}~\bibnamefont {Gebhardt}}, \bibinfo {author} {\bibfnamefont
  {B.}~\bibnamefont {Buda}}, \bibinfo {author} {\bibfnamefont {D.}~\bibnamefont
  {As}}, \bibinfo {author} {\bibfnamefont {D.}~\bibnamefont {Schikora}}, \emph
  {et~al.},\ }\bibfield  {title} {\bibinfo {title} {Temperature dependence of
  exciton peak energies in {ZnS, ZnSe, and ZnTe} epitaxial films},\ }\href@noop
  {} {\bibfield  {journal} {\bibinfo  {journal} {J. Appl. Phys.}\ }\textbf
  {\bibinfo {volume} {86}},\ \bibinfo {pages} {4403} (\bibinfo {year}
  {1999})}\BibitemShut {NoStop}%
\bibitem [{\citenamefont {Maruska}\ and\ \citenamefont
  {Tietjen}(1969)}]{maruska1969preparation}%
  \BibitemOpen
  \bibfield  {author} {\bibinfo {author} {\bibfnamefont {H.~P.}\ \bibnamefont
  {Maruska}}\ and\ \bibinfo {author} {\bibfnamefont {J.}~\bibnamefont
  {Tietjen}},\ }\bibfield  {title} {\bibinfo {title} {The preparation and
  properties of vapor-deposited single-crystal-line {GaN}},\ }\href
  {https://doi.org/10.1063/1.1652845} {\bibfield  {journal} {\bibinfo
  {journal} {Appl. Phys. Lett.}\ }\textbf {\bibinfo {volume} {15}},\ \bibinfo
  {pages} {327} (\bibinfo {year} {1969})}\BibitemShut {NoStop}%
\bibitem [{\citenamefont {Freitas~Jr}\ \emph {et~al.}(2001)\citenamefont
  {Freitas~Jr}, \citenamefont {Braga}, \citenamefont {Moore}, \citenamefont
  {Tischler}, \citenamefont {Culbertson}, \citenamefont {Fatemi}, \citenamefont
  {Park}, \citenamefont {Lee},\ and\ \citenamefont
  {Park}}]{freitas2001structural}%
  \BibitemOpen
  \bibfield  {author} {\bibinfo {author} {\bibfnamefont {J.}~\bibnamefont
  {Freitas~Jr}}, \bibinfo {author} {\bibfnamefont {G.}~\bibnamefont {Braga}},
  \bibinfo {author} {\bibfnamefont {W.}~\bibnamefont {Moore}}, \bibinfo
  {author} {\bibfnamefont {J.}~\bibnamefont {Tischler}}, \bibinfo {author}
  {\bibfnamefont {J.}~\bibnamefont {Culbertson}}, \bibinfo {author}
  {\bibfnamefont {M.}~\bibnamefont {Fatemi}}, \bibinfo {author} {\bibfnamefont
  {S.}~\bibnamefont {Park}}, \bibinfo {author} {\bibfnamefont {S.}~\bibnamefont
  {Lee}},\ and\ \bibinfo {author} {\bibfnamefont {Y.}~\bibnamefont {Park}},\
  }\bibfield  {title} {\bibinfo {title} {Structural and optical properties of
  thick freestanding {GaN} templates},\ }\href
  {https://doi.org/10.1016/S0022-0248(01)01456-7} {\bibfield  {journal}
  {\bibinfo  {journal} {J. Cryst. Growth}\ }\textbf {\bibinfo {volume} {231}},\
  \bibinfo {pages} {322} (\bibinfo {year} {2001})}\BibitemShut {NoStop}%
\bibitem [{\citenamefont {{\AA}hman}\ \emph {et~al.}(1996)\citenamefont
  {{\AA}hman}, \citenamefont {Svensson},\ and\ \citenamefont
  {Albertsson}}]{aahman1996reinvestigation}%
  \BibitemOpen
  \bibfield  {author} {\bibinfo {author} {\bibfnamefont {J.}~\bibnamefont
  {{\AA}hman}}, \bibinfo {author} {\bibfnamefont {G.}~\bibnamefont
  {Svensson}},\ and\ \bibinfo {author} {\bibfnamefont {J.}~\bibnamefont
  {Albertsson}},\ }\bibfield  {title} {\bibinfo {title} {A reinvestigation of
  $\beta$-gallium oxide},\ }\href {https://doi.org/10.1107/S0108270195016404}
  {\bibfield  {journal} {\bibinfo  {journal} {Acta Crystallogr., Sect. C:
  Cryst. Struct. Commun}\ }\textbf {\bibinfo {volume} {52}},\ \bibinfo {pages}
  {1336} (\bibinfo {year} {1996})}\BibitemShut {NoStop}%
\bibitem [{\citenamefont {Onuma}\ \emph {et~al.}(2015)\citenamefont {Onuma},
  \citenamefont {Saito}, \citenamefont {Sasaki}, \citenamefont {Masui},
  \citenamefont {Yamaguchi}, \citenamefont {Honda},\ and\ \citenamefont
  {Higashiwaki}}]{onuma2015valence}%
  \BibitemOpen
  \bibfield  {author} {\bibinfo {author} {\bibfnamefont {T.}~\bibnamefont
  {Onuma}}, \bibinfo {author} {\bibfnamefont {S.}~\bibnamefont {Saito}},
  \bibinfo {author} {\bibfnamefont {K.}~\bibnamefont {Sasaki}}, \bibinfo
  {author} {\bibfnamefont {T.}~\bibnamefont {Masui}}, \bibinfo {author}
  {\bibfnamefont {T.}~\bibnamefont {Yamaguchi}}, \bibinfo {author}
  {\bibfnamefont {T.}~\bibnamefont {Honda}},\ and\ \bibinfo {author}
  {\bibfnamefont {M.}~\bibnamefont {Higashiwaki}},\ }\bibfield  {title}
  {\bibinfo {title} {Valence band ordering in $\beta$-{Ga$_2$O$_3$} studied by
  polarized transmittance and reflectance spectroscopy},\ }\href
  {https://iopscience.iop.org/article/10.7567/JJAP.54.112601/meta} {\bibfield
  {journal} {\bibinfo  {journal} {Jpn. J. Appl. Phys.}\ }\textbf {\bibinfo
  {volume} {54}},\ \bibinfo {pages} {112601} (\bibinfo {year}
  {2015})}\BibitemShut {NoStop}%
\bibitem [{\citenamefont {Prencipe}\ \emph {et~al.}(1995)\citenamefont
  {Prencipe}, \citenamefont {Zupan}, \citenamefont {Dovesi}, \citenamefont
  {Apra},\ and\ \citenamefont {Saunders}}]{prencipe1995ab}%
  \BibitemOpen
  \bibfield  {author} {\bibinfo {author} {\bibfnamefont {M.}~\bibnamefont
  {Prencipe}}, \bibinfo {author} {\bibfnamefont {A.}~\bibnamefont {Zupan}},
  \bibinfo {author} {\bibfnamefont {R.}~\bibnamefont {Dovesi}}, \bibinfo
  {author} {\bibfnamefont {E.}~\bibnamefont {Apra}},\ and\ \bibinfo {author}
  {\bibfnamefont {V.}~\bibnamefont {Saunders}},\ }\bibfield  {title} {\bibinfo
  {title} {Ab initio study of the structural properties of {LiF, NaF, KF, LiCl,
  NaCl, and KCl}},\ }\href {https://doi.org/10.1103/PhysRevB.51.3391}
  {\bibfield  {journal} {\bibinfo  {journal} {Phys. Rev. B}\ }\textbf {\bibinfo
  {volume} {51}},\ \bibinfo {pages} {3391} (\bibinfo {year}
  {1995})}\BibitemShut {NoStop}%
\bibitem [{\citenamefont {Alkauskas}\ \emph {et~al.}(2008)\citenamefont
  {Alkauskas}, \citenamefont {Broqvist},\ and\ \citenamefont
  {Pasquarello}}]{alkauskas2008defect}%
  \BibitemOpen
  \bibfield  {author} {\bibinfo {author} {\bibfnamefont {A.}~\bibnamefont
  {Alkauskas}}, \bibinfo {author} {\bibfnamefont {P.}~\bibnamefont
  {Broqvist}},\ and\ \bibinfo {author} {\bibfnamefont {A.}~\bibnamefont
  {Pasquarello}},\ }\bibfield  {title} {\bibinfo {title} {Defect energy levels
  in density functional calculations: Alignment and band gap problem},\ }\href
  {https://doi.org/10.1103/PhysRevLett.101.046405} {\bibfield  {journal}
  {\bibinfo  {journal} {Phys. Rev. Lett.}\ }\textbf {\bibinfo {volume} {101}},\
  \bibinfo {pages} {046405} (\bibinfo {year} {2008})}\BibitemShut {NoStop}%
\bibitem [{\citenamefont {Frodason}\ \emph {et~al.}(2017)\citenamefont
  {Frodason}, \citenamefont {Johansen}, \citenamefont {Bj{\o}rheim},
  \citenamefont {Svensson},\ and\ \citenamefont {Alkauskas}}]{frodason2017zn}%
  \BibitemOpen
  \bibfield  {author} {\bibinfo {author} {\bibfnamefont {Y.}~\bibnamefont
  {Frodason}}, \bibinfo {author} {\bibfnamefont {K.}~\bibnamefont {Johansen}},
  \bibinfo {author} {\bibfnamefont {T.}~\bibnamefont {Bj{\o}rheim}}, \bibinfo
  {author} {\bibfnamefont {B.}~\bibnamefont {Svensson}},\ and\ \bibinfo
  {author} {\bibfnamefont {A.}~\bibnamefont {Alkauskas}},\ }\bibfield  {title}
  {\bibinfo {title} {Zn vacancy as a polaronic hole trap in {ZnO}},\ }\href
  {https://doi.org/10.1103/PhysRevB.95.094105} {\bibfield  {journal} {\bibinfo
  {journal} {Phys. Rev. B}\ }\textbf {\bibinfo {volume} {95}},\ \bibinfo
  {pages} {094105} (\bibinfo {year} {2017})}\BibitemShut {NoStop}%
\bibitem [{\citenamefont {Lyons}\ \emph {et~al.}(2017)\citenamefont {Lyons},
  \citenamefont {Varley}, \citenamefont {Steiauf}, \citenamefont {Janotti},\
  and\ \citenamefont {Van~de Walle}}]{Lyons17}%
  \BibitemOpen
  \bibfield  {author} {\bibinfo {author} {\bibfnamefont {J.~L.}\ \bibnamefont
  {Lyons}}, \bibinfo {author} {\bibfnamefont {J.~B.}\ \bibnamefont {Varley}},
  \bibinfo {author} {\bibfnamefont {D.}~\bibnamefont {Steiauf}}, \bibinfo
  {author} {\bibfnamefont {A.}~\bibnamefont {Janotti}},\ and\ \bibinfo {author}
  {\bibfnamefont {C.~G.}\ \bibnamefont {Van~de Walle}},\ }\bibfield  {title}
  {\bibinfo {title} {First-principles characterization of native-defect-related
  optical transitions in {ZnO}},\ }\href {https://doi.org/10.1063/1.4992128}
  {\bibfield  {journal} {\bibinfo  {journal} {J. Appl. Phys.}\ }\textbf
  {\bibinfo {volume} {122}},\ \bibinfo {pages} {035704} (\bibinfo {year}
  {2017})}\BibitemShut {NoStop}%
\bibitem [{\citenamefont {Evans}\ \emph {et~al.}(2008)\citenamefont {Evans},
  \citenamefont {Giles}, \citenamefont {Halliburton},\ and\ \citenamefont
  {Kappers}}]{evans2008further}%
  \BibitemOpen
  \bibfield  {author} {\bibinfo {author} {\bibfnamefont {S.}~\bibnamefont
  {Evans}}, \bibinfo {author} {\bibfnamefont {N.}~\bibnamefont {Giles}},
  \bibinfo {author} {\bibfnamefont {L.}~\bibnamefont {Halliburton}},\ and\
  \bibinfo {author} {\bibfnamefont {L.}~\bibnamefont {Kappers}},\ }\bibfield
  {title} {\bibinfo {title} {Further characterization of oxygen vacancies and
  zinc vacancies in electron-irradiated {ZnO}},\ }\href
  {https://doi.org/10.1063/1.2833432} {\bibfield  {journal} {\bibinfo
  {journal} {J. Appl. Phys.}\ }\textbf {\bibinfo {volume} {103}},\ \bibinfo
  {pages} {043710} (\bibinfo {year} {2008})}\BibitemShut {NoStop}%
\bibitem [{\citenamefont {Oba}\ \emph {et~al.}(2008)\citenamefont {Oba},
  \citenamefont {Togo}, \citenamefont {Tanaka}, \citenamefont {Paier},\ and\
  \citenamefont {Kresse}}]{oba2008defect}%
  \BibitemOpen
  \bibfield  {author} {\bibinfo {author} {\bibfnamefont {F.}~\bibnamefont
  {Oba}}, \bibinfo {author} {\bibfnamefont {A.}~\bibnamefont {Togo}}, \bibinfo
  {author} {\bibfnamefont {I.}~\bibnamefont {Tanaka}}, \bibinfo {author}
  {\bibfnamefont {J.}~\bibnamefont {Paier}},\ and\ \bibinfo {author}
  {\bibfnamefont {G.}~\bibnamefont {Kresse}},\ }\bibfield  {title} {\bibinfo
  {title} {Defect energetics in {ZnO}: A hybrid hartree-fock density functional
  study},\ }\href@noop {} {\bibfield  {journal} {\bibinfo  {journal} {Phys.
  Rev. B}\ }\textbf {\bibinfo {volume} {77}},\ \bibinfo {pages} {245202}
  (\bibinfo {year} {2008})}\BibitemShut {NoStop}%
\bibitem [{\citenamefont {Alkauskas}\ and\ \citenamefont
  {Pasquarello}(2011)}]{alkauskas2011band}%
  \BibitemOpen
  \bibfield  {author} {\bibinfo {author} {\bibfnamefont {A.}~\bibnamefont
  {Alkauskas}}\ and\ \bibinfo {author} {\bibfnamefont {A.}~\bibnamefont
  {Pasquarello}},\ }\bibfield  {title} {\bibinfo {title} {Band-edge problem in
  the theoretical determination of defect energy levels: The o vacancy in zno
  as a benchmark case},\ }\href@noop {} {\bibfield  {journal} {\bibinfo
  {journal} {Phys. Rev. B}\ }\textbf {\bibinfo {volume} {84}},\ \bibinfo
  {pages} {125206} (\bibinfo {year} {2011})}\BibitemShut {NoStop}%
\bibitem [{\citenamefont {Jeon}\ \emph {et~al.}(1993)\citenamefont {Jeon},
  \citenamefont {Gislason},\ and\ \citenamefont {Watkins}}]{jeon1993optical}%
  \BibitemOpen
  \bibfield  {author} {\bibinfo {author} {\bibfnamefont {D.~Y.}\ \bibnamefont
  {Jeon}}, \bibinfo {author} {\bibfnamefont {H.}~\bibnamefont {Gislason}},\
  and\ \bibinfo {author} {\bibfnamefont {G.~D.}\ \bibnamefont {Watkins}},\
  }\bibfield  {title} {\bibinfo {title} {Optical detection of magnetic
  resonance of the zinc vacancy in znse via magnetic circular dichroism},\
  }\href@noop {} {\bibfield  {journal} {\bibinfo  {journal} {Phys. Rev. B}\
  }\textbf {\bibinfo {volume} {48}},\ \bibinfo {pages} {7872} (\bibinfo {year}
  {1993})}\BibitemShut {NoStop}%
\bibitem [{\citenamefont {Shen}\ \emph {et~al.}(2017)\citenamefont {Shen},
  \citenamefont {Wickramaratne}, \citenamefont {Dreyer}, \citenamefont
  {Alkauskas}, \citenamefont {Young}, \citenamefont {Speck},\ and\
  \citenamefont {Van~de Walle}}]{shen2017calcium}%
  \BibitemOpen
  \bibfield  {author} {\bibinfo {author} {\bibfnamefont {J.-X.}\ \bibnamefont
  {Shen}}, \bibinfo {author} {\bibfnamefont {D.}~\bibnamefont {Wickramaratne}},
  \bibinfo {author} {\bibfnamefont {C.~E.}\ \bibnamefont {Dreyer}}, \bibinfo
  {author} {\bibfnamefont {A.}~\bibnamefont {Alkauskas}}, \bibinfo {author}
  {\bibfnamefont {E.}~\bibnamefont {Young}}, \bibinfo {author} {\bibfnamefont
  {J.~S.}\ \bibnamefont {Speck}},\ and\ \bibinfo {author} {\bibfnamefont
  {C.~G.}\ \bibnamefont {Van~de Walle}},\ }\bibfield  {title} {\bibinfo {title}
  {Calcium as a nonradiative recombination center in {InGaN}},\ }\href
  {https://iopscience.iop.org/article/10.7567/APEX.10.021001/meta} {\bibfield
  {journal} {\bibinfo  {journal} {Appl. Phys. Express}\ }\textbf {\bibinfo
  {volume} {10}},\ \bibinfo {pages} {021001} (\bibinfo {year}
  {2017})}\BibitemShut {NoStop}%
\bibitem [{\citenamefont {Young}\ \emph {et~al.}(2016)\citenamefont {Young},
  \citenamefont {Grandjean}, \citenamefont {Mates},\ and\ \citenamefont
  {Speck}}]{young2016calcium}%
  \BibitemOpen
  \bibfield  {author} {\bibinfo {author} {\bibfnamefont {E.}~\bibnamefont
  {Young}}, \bibinfo {author} {\bibfnamefont {N.}~\bibnamefont {Grandjean}},
  \bibinfo {author} {\bibfnamefont {T.}~\bibnamefont {Mates}},\ and\ \bibinfo
  {author} {\bibfnamefont {J.}~\bibnamefont {Speck}},\ }\bibfield  {title}
  {\bibinfo {title} {Calcium impurity as a source of non-radiative
  recombination in {(In, Ga) N} layers grown by molecular beam epitaxy},\
  }\href {https://doi.org/10.1063/1.4968586} {\bibfield  {journal} {\bibinfo
  {journal} {Appl. Phys. Lett.}\ }\textbf {\bibinfo {volume} {109}},\ \bibinfo
  {pages} {212103} (\bibinfo {year} {2016})}\BibitemShut {NoStop}%
\bibitem [{\citenamefont {Zhao}\ \emph {et~al.}(2023)\citenamefont {Zhao},
  \citenamefont {Turiansky}, \citenamefont {Alkauskas},\ and\ \citenamefont
  {Van~de Walle}}]{zhao2023trap}%
  \BibitemOpen
  \bibfield  {author} {\bibinfo {author} {\bibfnamefont {F.}~\bibnamefont
  {Zhao}}, \bibinfo {author} {\bibfnamefont {M.~E.}\ \bibnamefont {Turiansky}},
  \bibinfo {author} {\bibfnamefont {A.}~\bibnamefont {Alkauskas}},\ and\
  \bibinfo {author} {\bibfnamefont {C.~G.}\ \bibnamefont {Van~de Walle}},\
  }\bibfield  {title} {\bibinfo {title} {Trap-assisted {Auger-Meitner}
  recombination from first principles},\ }\href
  {https://doi.org/10.1103/PhysRevLett.131.056402} {\bibfield  {journal}
  {\bibinfo  {journal} {Phys. Rev. Lett.}\ }\textbf {\bibinfo {volume} {131}},\
  \bibinfo {pages} {056402} (\bibinfo {year} {2023})}\BibitemShut {NoStop}%
\bibitem [{\citenamefont {Lyons}\ \emph {et~al.}(2013)\citenamefont {Lyons},
  \citenamefont {Janotti},\ and\ \citenamefont {Van~de
  Walle}}]{lyons2013impact}%
  \BibitemOpen
  \bibfield  {author} {\bibinfo {author} {\bibfnamefont {J.~L.}\ \bibnamefont
  {Lyons}}, \bibinfo {author} {\bibfnamefont {A.}~\bibnamefont {Janotti}},\
  and\ \bibinfo {author} {\bibfnamefont {C.~G.}\ \bibnamefont {Van~de Walle}},\
  }\bibfield  {title} {\bibinfo {title} {Impact of group-{II} acceptors on the
  electrical and optical properties of {GaN}},\ }\href
  {https://iopscience.iop.org/article/10.7567/JJAP.52.08JJ04/meta} {\bibfield
  {journal} {\bibinfo  {journal} {Japanese J. Appl. Phys.}\ }\textbf {\bibinfo
  {volume} {52}},\ \bibinfo {pages} {08JJ04} (\bibinfo {year}
  {2013})}\BibitemShut {NoStop}%
\bibitem [{\citenamefont {Kanegae}\ \emph {et~al.}(2021)\citenamefont
  {Kanegae}, \citenamefont {Narita}, \citenamefont {Tomita}, \citenamefont
  {Kachi}, \citenamefont {Horita}, \citenamefont {Kimoto},\ and\ \citenamefont
  {Suda}}]{kanegae2021photoionization}%
  \BibitemOpen
  \bibfield  {author} {\bibinfo {author} {\bibfnamefont {K.}~\bibnamefont
  {Kanegae}}, \bibinfo {author} {\bibfnamefont {T.}~\bibnamefont {Narita}},
  \bibinfo {author} {\bibfnamefont {K.}~\bibnamefont {Tomita}}, \bibinfo
  {author} {\bibfnamefont {T.}~\bibnamefont {Kachi}}, \bibinfo {author}
  {\bibfnamefont {M.}~\bibnamefont {Horita}}, \bibinfo {author} {\bibfnamefont
  {T.}~\bibnamefont {Kimoto}},\ and\ \bibinfo {author} {\bibfnamefont
  {J.}~\bibnamefont {Suda}},\ }\bibfield  {title} {\bibinfo {title}
  {Photoionization cross section ratio of nitrogen-site carbon in {GaN} under
  sub-bandgap-light irradiation determined by isothermal capacitance transient
  spectroscopy},\ }\href
  {https://iopscience.iop.org/article/10.35848/1882-0786/ac16ba/meta}
  {\bibfield  {journal} {\bibinfo  {journal} {Appl. Phys. Express}\ }\textbf
  {\bibinfo {volume} {14}},\ \bibinfo {pages} {091004} (\bibinfo {year}
  {2021})}\BibitemShut {NoStop}%
\bibitem [{\citenamefont {Lyons}\ \emph {et~al.}(2010)\citenamefont {Lyons},
  \citenamefont {Janotti},\ and\ \citenamefont {{Van de
  Walle}}}]{lyons2010carbon}%
  \BibitemOpen
  \bibfield  {author} {\bibinfo {author} {\bibfnamefont {J.~L.}\ \bibnamefont
  {Lyons}}, \bibinfo {author} {\bibfnamefont {A.}~\bibnamefont {Janotti}},\
  and\ \bibinfo {author} {\bibfnamefont {C.~G.}\ \bibnamefont {{Van de
  Walle}}},\ }\bibfield  {title} {\bibinfo {title} {Carbon impurities and the
  yellow luminescence in {GaN}},\ }\href {https://doi.org/10.1063/1.3492841}
  {\bibfield  {journal} {\bibinfo  {journal} {Appl. Phys. Lett.}\ }\textbf
  {\bibinfo {volume} {97}},\ \bibinfo {pages} {152108} (\bibinfo {year}
  {2010})}\BibitemShut {NoStop}%
\bibitem [{\citenamefont {Bhandari}\ \emph {et~al.}(2022)\citenamefont
  {Bhandari}, \citenamefont {Lyons}, \citenamefont {Wickramaratne},\ and\
  \citenamefont {Zvanut}}]{bhandari2022optical}%
  \BibitemOpen
  \bibfield  {author} {\bibinfo {author} {\bibfnamefont {S.}~\bibnamefont
  {Bhandari}}, \bibinfo {author} {\bibfnamefont {J.~L.}\ \bibnamefont {Lyons}},
  \bibinfo {author} {\bibfnamefont {D.}~\bibnamefont {Wickramaratne}},\ and\
  \bibinfo {author} {\bibfnamefont {M.~E.}\ \bibnamefont {Zvanut}},\ }\bibfield
   {title} {\bibinfo {title} {Optical transitions of neutral {Mg} in {Mg}-doped
  $\beta$-{Ga$_2$O$_3$}},\ }\href {https://doi.org/10.1063/5.0081925}
  {\bibfield  {journal} {\bibinfo  {journal} {APL Mater.}\ }\textbf {\bibinfo
  {volume} {10}},\ \bibinfo {pages} {021103} (\bibinfo {year}
  {2022})}\BibitemShut {NoStop}%
\bibitem [{\citenamefont {Kananen}\ \emph {et~al.}(2017)\citenamefont
  {Kananen}, \citenamefont {Halliburton}, \citenamefont {Scherrer},
  \citenamefont {Stevens}, \citenamefont {Foundos}, \citenamefont {Chang},\
  and\ \citenamefont {Giles}}]{kananen2017electron}%
  \BibitemOpen
  \bibfield  {author} {\bibinfo {author} {\bibfnamefont {B.~E.}\ \bibnamefont
  {Kananen}}, \bibinfo {author} {\bibfnamefont {L.~E.}\ \bibnamefont
  {Halliburton}}, \bibinfo {author} {\bibfnamefont {E.~M.}\ \bibnamefont
  {Scherrer}}, \bibinfo {author} {\bibfnamefont {K.}~\bibnamefont {Stevens}},
  \bibinfo {author} {\bibfnamefont {G.}~\bibnamefont {Foundos}}, \bibinfo
  {author} {\bibfnamefont {K.}~\bibnamefont {Chang}},\ and\ \bibinfo {author}
  {\bibfnamefont {N.~C.}\ \bibnamefont {Giles}},\ }\bibfield  {title} {\bibinfo
  {title} {Electron paramagnetic resonance study of neutral mg acceptors in
  $\beta$-{Ga$_2$O$_3$} crystals},\ }\href {https://doi.org/10.1063/1.4990454}
  {\bibfield  {journal} {\bibinfo  {journal} {Appl. Phys. Lett.}\ }\textbf
  {\bibinfo {volume} {111}},\ \bibinfo {pages} {072102} (\bibinfo {year}
  {2017})}\BibitemShut {NoStop}%
\bibitem [{\citenamefont {Varley}\ \emph {et~al.}(2012)\citenamefont {Varley},
  \citenamefont {Janotti}, \citenamefont {Franchini},\ and\ \citenamefont
  {Van~de Walle}}]{varley2012role}%
  \BibitemOpen
  \bibfield  {author} {\bibinfo {author} {\bibfnamefont {J.~B.}\ \bibnamefont
  {Varley}}, \bibinfo {author} {\bibfnamefont {A.}~\bibnamefont {Janotti}},
  \bibinfo {author} {\bibfnamefont {C.}~\bibnamefont {Franchini}},\ and\
  \bibinfo {author} {\bibfnamefont {C.~G.}\ \bibnamefont {Van~de Walle}},\
  }\bibfield  {title} {\bibinfo {title} {Role of self-trapping in luminescence
  and p-type conductivity of wide-band-gap oxides},\ }\href
  {https://doi.org/10.1103/PhysRevB.85.081109} {\bibfield  {journal} {\bibinfo
  {journal} {Phys. Rev. B}\ }\textbf {\bibinfo {volume} {85}},\ \bibinfo
  {pages} {081109} (\bibinfo {year} {2012})}\BibitemShut {NoStop}%
\bibitem [{\citenamefont {Lyons}(2022)}]{lyons2022self}%
  \BibitemOpen
  \bibfield  {author} {\bibinfo {author} {\bibfnamefont {J.~L.}\ \bibnamefont
  {Lyons}},\ }\bibfield  {title} {\bibinfo {title} {Self-trapped holes and
  polaronic acceptors in ultrawide-bandgap oxides},\ }\href
  {https://doi.org/10.1063/5.0077030} {\bibfield  {journal} {\bibinfo
  {journal} {J. Appl. Phys.}\ }\textbf {\bibinfo {volume} {131}},\ \bibinfo
  {pages} {025701} (\bibinfo {year} {2022})}\BibitemShut {NoStop}%
\bibitem [{\citenamefont {Williams}\ and\ \citenamefont
  {Song}(1990)}]{williams1990self}%
  \BibitemOpen
  \bibfield  {author} {\bibinfo {author} {\bibfnamefont {R.}~\bibnamefont
  {Williams}}\ and\ \bibinfo {author} {\bibfnamefont {K.}~\bibnamefont
  {Song}},\ }\bibfield  {title} {\bibinfo {title} {The self-trapped exciton},\
  }\href {https://doi.org/10.1016/0022-3697(90)90144-5} {\bibfield  {journal}
  {\bibinfo  {journal} {J. Phys. Chem. Solids}\ }\textbf {\bibinfo {volume}
  {51}},\ \bibinfo {pages} {679} (\bibinfo {year} {1990})}\BibitemShut
  {NoStop}%
\bibitem [{\citenamefont {Miceli}\ \emph {et~al.}(2018)\citenamefont {Miceli},
  \citenamefont {Chen}, \citenamefont {Reshetnyak},\ and\ \citenamefont
  {Pasquarello}}]{miceli2018nonempirical}%
  \BibitemOpen
  \bibfield  {author} {\bibinfo {author} {\bibfnamefont {G.}~\bibnamefont
  {Miceli}}, \bibinfo {author} {\bibfnamefont {W.}~\bibnamefont {Chen}},
  \bibinfo {author} {\bibfnamefont {I.}~\bibnamefont {Reshetnyak}},\ and\
  \bibinfo {author} {\bibfnamefont {A.}~\bibnamefont {Pasquarello}},\
  }\bibfield  {title} {\bibinfo {title} {Nonempirical hybrid functionals for
  band gaps and polaronic distortions in solids},\ }\href@noop {} {\bibfield
  {journal} {\bibinfo  {journal} {Phys. Rev. B}\ }\textbf {\bibinfo {volume}
  {97}},\ \bibinfo {pages} {121112} (\bibinfo {year} {2018})}\BibitemShut
  {NoStop}%
\bibitem [{\citenamefont {Turiansky}\ \emph {et~al.}(2021)\citenamefont
  {Turiansky}, \citenamefont {Alkauskas}, \citenamefont {Engel}, \citenamefont
  {Kresse}, \citenamefont {Wickramaratne}, \citenamefont {Shen}, \citenamefont
  {Dreyer},\ and\ \citenamefont {Van~de Walle}}]{turiansky2021nonrad}%
  \BibitemOpen
  \bibfield  {author} {\bibinfo {author} {\bibfnamefont {M.~E.}\ \bibnamefont
  {Turiansky}}, \bibinfo {author} {\bibfnamefont {A.}~\bibnamefont
  {Alkauskas}}, \bibinfo {author} {\bibfnamefont {M.}~\bibnamefont {Engel}},
  \bibinfo {author} {\bibfnamefont {G.}~\bibnamefont {Kresse}}, \bibinfo
  {author} {\bibfnamefont {D.}~\bibnamefont {Wickramaratne}}, \bibinfo {author}
  {\bibfnamefont {J.-X.}\ \bibnamefont {Shen}}, \bibinfo {author}
  {\bibfnamefont {C.~E.}\ \bibnamefont {Dreyer}},\ and\ \bibinfo {author}
  {\bibfnamefont {C.~G.}\ \bibnamefont {Van~de Walle}},\ }\bibfield  {title}
  {\bibinfo {title} {Nonrad: Computing nonradiative capture coefficients from
  first principles},\ }\href {https://doi.org/10.1016/j.cpc.2021.108056}
  {\bibfield  {journal} {\bibinfo  {journal} {Comput. Phys. Commun.}\ }\textbf
  {\bibinfo {volume} {267}},\ \bibinfo {pages} {108056} (\bibinfo {year}
  {2021})}\BibitemShut {NoStop}%
\bibitem [{\citenamefont {Dreyer}\ \emph {et~al.}(2020)\citenamefont {Dreyer},
  \citenamefont {Alkauskas}, \citenamefont {Lyons},\ and\ \citenamefont {Van~de
  Walle}}]{dreyer2020radiative}%
  \BibitemOpen
  \bibfield  {author} {\bibinfo {author} {\bibfnamefont {C.~E.}\ \bibnamefont
  {Dreyer}}, \bibinfo {author} {\bibfnamefont {A.}~\bibnamefont {Alkauskas}},
  \bibinfo {author} {\bibfnamefont {J.~L.}\ \bibnamefont {Lyons}},\ and\
  \bibinfo {author} {\bibfnamefont {C.~G.}\ \bibnamefont {Van~de Walle}},\
  }\bibfield  {title} {\bibinfo {title} {Radiative capture rates at deep
  defects from electronic structure calculations},\ }\href
  {https://doi.org/10.1103/PhysRevB.102.085305} {\bibfield  {journal} {\bibinfo
   {journal} {Phys. Rev. B}\ }\textbf {\bibinfo {volume} {102}},\ \bibinfo
  {pages} {085305} (\bibinfo {year} {2020})}\BibitemShut {NoStop}%
\bibitem [{\citenamefont {Alkauskas}\ \emph {et~al.}(2016)\citenamefont
  {Alkauskas}, \citenamefont {McCluskey},\ and\ \citenamefont {Van~de
  Walle}}]{alkauskas2016tutorial}%
  \BibitemOpen
  \bibfield  {author} {\bibinfo {author} {\bibfnamefont {A.}~\bibnamefont
  {Alkauskas}}, \bibinfo {author} {\bibfnamefont {M.~D.}\ \bibnamefont
  {McCluskey}},\ and\ \bibinfo {author} {\bibfnamefont {C.~G.}\ \bibnamefont
  {Van~de Walle}},\ }\bibfield  {title} {\bibinfo {title} {Tutorial: Defects in
  semiconductors—combining experiment and theory},\ }\href
  {https://doi.org/10.1063/1.4948245} {\bibfield  {journal} {\bibinfo
  {journal} {J. Appl. Phys.}\ }\textbf {\bibinfo {volume} {119}},\ \bibinfo
  {pages} {181101} (\bibinfo {year} {2016})}\BibitemShut {NoStop}%
\bibitem [{\citenamefont {Wang}\ \emph {et~al.}(2009)\citenamefont {Wang},
  \citenamefont {Vlasenko}, \citenamefont {Pearton}, \citenamefont {Chen},\
  and\ \citenamefont {Buyanova}}]{wang2009oxygen}%
  \BibitemOpen
  \bibfield  {author} {\bibinfo {author} {\bibfnamefont {X.}~\bibnamefont
  {Wang}}, \bibinfo {author} {\bibfnamefont {L.}~\bibnamefont {Vlasenko}},
  \bibinfo {author} {\bibfnamefont {S.}~\bibnamefont {Pearton}}, \bibinfo
  {author} {\bibfnamefont {W.}~\bibnamefont {Chen}},\ and\ \bibinfo {author}
  {\bibfnamefont {I.~A.}\ \bibnamefont {Buyanova}},\ }\bibfield  {title}
  {\bibinfo {title} {Oxygen and zinc vacancies in as-grown zno single
  crystals},\ }\href@noop {} {\bibfield  {journal} {\bibinfo  {journal}
  {Journal of Physics D: Applied Physics}\ }\textbf {\bibinfo {volume} {42}},\
  \bibinfo {pages} {175411} (\bibinfo {year} {2009})}\BibitemShut {NoStop}%
\bibitem [{\citenamefont {Ogino}\ and\ \citenamefont
  {Aoki}(1980)}]{ogino1980mechanism}%
  \BibitemOpen
  \bibfield  {author} {\bibinfo {author} {\bibfnamefont {T.}~\bibnamefont
  {Ogino}}\ and\ \bibinfo {author} {\bibfnamefont {M.}~\bibnamefont {Aoki}},\
  }\bibfield  {title} {\bibinfo {title} {Mechanism of yellow luminescence in
  {GaN}},\ }\href
  {https://iopscience.iop.org/article/10.1143/JJAP.19.2395/meta} {\bibfield
  {journal} {\bibinfo  {journal} {Japanese J. Appl. Phys.}\ }\textbf {\bibinfo
  {volume} {19}},\ \bibinfo {pages} {2395} (\bibinfo {year}
  {1980})}\BibitemShut {NoStop}%
\bibitem [{\citenamefont {Falletta}\ \emph {et~al.}(2020)\citenamefont
  {Falletta}, \citenamefont {Wiktor},\ and\ \citenamefont
  {Pasquarello}}]{falletta2020finite}%
  \BibitemOpen
  \bibfield  {author} {\bibinfo {author} {\bibfnamefont {S.}~\bibnamefont
  {Falletta}}, \bibinfo {author} {\bibfnamefont {J.}~\bibnamefont {Wiktor}},\
  and\ \bibinfo {author} {\bibfnamefont {A.}~\bibnamefont {Pasquarello}},\
  }\bibfield  {title} {\bibinfo {title} {Finite-size corrections of defect
  energy levels involving ionic polarization},\ }\href@noop {} {\bibfield
  {journal} {\bibinfo  {journal} {Phys. Rev. B}\ }\textbf {\bibinfo {volume}
  {102}},\ \bibinfo {pages} {041115} (\bibinfo {year} {2020})}\BibitemShut
  {NoStop}%
\bibitem [{\citenamefont {Gake}\ \emph {et~al.}(2020)\citenamefont {Gake},
  \citenamefont {Kumagai}, \citenamefont {Freysoldt},\ and\ \citenamefont
  {Oba}}]{gake2020finite}%
  \BibitemOpen
  \bibfield  {author} {\bibinfo {author} {\bibfnamefont {T.}~\bibnamefont
  {Gake}}, \bibinfo {author} {\bibfnamefont {Y.}~\bibnamefont {Kumagai}},
  \bibinfo {author} {\bibfnamefont {C.}~\bibnamefont {Freysoldt}},\ and\
  \bibinfo {author} {\bibfnamefont {F.}~\bibnamefont {Oba}},\ }\bibfield
  {title} {\bibinfo {title} {Finite-size corrections for defect-involving
  vertical transitions in supercell calculations},\ }\href@noop {} {\bibfield
  {journal} {\bibinfo  {journal} {Phys. Rev. B}\ }\textbf {\bibinfo {volume}
  {101}},\ \bibinfo {pages} {020102} (\bibinfo {year} {2020})}\BibitemShut
  {NoStop}%
\bibitem [{\citenamefont {Kumagai}(2023)}]{kumagai2023finite}%
  \BibitemOpen
  \bibfield  {author} {\bibinfo {author} {\bibfnamefont {Y.}~\bibnamefont
  {Kumagai}},\ }\bibfield  {title} {\bibinfo {title} {Finite-size corrections
  to defect energetics along one-dimensional configuration coordinate},\
  }\href@noop {} {\bibfield  {journal} {\bibinfo  {journal} {Phys. Rev. B}\
  }\textbf {\bibinfo {volume} {107}},\ \bibinfo {pages} {L220101} (\bibinfo
  {year} {2023})}\BibitemShut {NoStop}%
\bibitem [{\citenamefont {Long}\ \emph {et~al.}(2020)\citenamefont {Long},
  \citenamefont {Gautam},\ and\ \citenamefont {Carter}}]{long2020evaluating}%
  \BibitemOpen
  \bibfield  {author} {\bibinfo {author} {\bibfnamefont {O.~Y.}\ \bibnamefont
  {Long}}, \bibinfo {author} {\bibfnamefont {G.~S.}\ \bibnamefont {Gautam}},\
  and\ \bibinfo {author} {\bibfnamefont {E.~A.}\ \bibnamefont {Carter}},\
  }\bibfield  {title} {\bibinfo {title} {Evaluating optimal {U} for 3d
  transition-metal oxides within the {SCAN+U} framework},\ }\href
  {https://doi.org/10.1103/PhysRevMaterials.4.045401} {\bibfield  {journal}
  {\bibinfo  {journal} {Phys. Rev. Mater.}\ }\textbf {\bibinfo {volume} {4}},\
  \bibinfo {pages} {045401} (\bibinfo {year} {2020})}\BibitemShut {NoStop}%
\bibitem [{\citenamefont {Lany}\ and\ \citenamefont
  {Zunger}(2009)}]{lany2009polaronic}%
  \BibitemOpen
  \bibfield  {author} {\bibinfo {author} {\bibfnamefont {S.}~\bibnamefont
  {Lany}}\ and\ \bibinfo {author} {\bibfnamefont {A.}~\bibnamefont {Zunger}},\
  }\bibfield  {title} {\bibinfo {title} {Polaronic hole localization and
  multiple hole binding of acceptors in oxide wide-gap semiconductors},\
  }\href@noop {} {\bibfield  {journal} {\bibinfo  {journal} {Phys. Rev. B}\
  }\textbf {\bibinfo {volume} {80}},\ \bibinfo {pages} {085202} (\bibinfo
  {year} {2009})}\BibitemShut {NoStop}%
\bibitem [{\citenamefont {Falletta}\ and\ \citenamefont
  {Pasquarello}(2022)}]{falletta2022polarons}%
  \BibitemOpen
  \bibfield  {author} {\bibinfo {author} {\bibfnamefont {S.}~\bibnamefont
  {Falletta}}\ and\ \bibinfo {author} {\bibfnamefont {A.}~\bibnamefont
  {Pasquarello}},\ }\bibfield  {title} {\bibinfo {title} {Polarons free from
  many-body self-interaction in density functional theory},\ }\href@noop {}
  {\bibfield  {journal} {\bibinfo  {journal} {Phys. Rev. B}\ }\textbf {\bibinfo
  {volume} {106}},\ \bibinfo {pages} {125119} (\bibinfo {year}
  {2022})}\BibitemShut {NoStop}%
\bibitem [{\citenamefont {Janotti}\ and\ \citenamefont {Van~de
  Walle}(2007)}]{janotti2007native}%
  \BibitemOpen
  \bibfield  {author} {\bibinfo {author} {\bibfnamefont {A.}~\bibnamefont
  {Janotti}}\ and\ \bibinfo {author} {\bibfnamefont {C.~G.}\ \bibnamefont
  {Van~de Walle}},\ }\bibfield  {title} {\bibinfo {title} {Native point defects
  in {ZnO}},\ }\href {https://doi.org/10.1103/PhysRevB.76.165202} {\bibfield
  {journal} {\bibinfo  {journal} {Phys. Rev. B}\ }\textbf {\bibinfo {volume}
  {76}},\ \bibinfo {pages} {165202} (\bibinfo {year} {2007})}\BibitemShut
  {NoStop}%
\bibitem [{\citenamefont {Hohenberg}\ and\ \citenamefont {Kohn}(1964)}]{HK64}%
  \BibitemOpen
  \bibfield  {author} {\bibinfo {author} {\bibfnamefont {P.}~\bibnamefont
  {Hohenberg}}\ and\ \bibinfo {author} {\bibfnamefont {W.}~\bibnamefont
  {Kohn}},\ }\bibfield  {title} {\bibinfo {title} {Inhomogeneous electron
  gas},\ }\href {https://doi.org/10.1103/PhysRev.136.B864} {\bibfield
  {journal} {\bibinfo  {journal} {Phys. Rev.}\ }\textbf {\bibinfo {volume}
  {136}},\ \bibinfo {pages} {B864} (\bibinfo {year} {1964})}\BibitemShut
  {NoStop}%
\bibitem [{\citenamefont {Kohn}\ and\ \citenamefont {Sham}(1965)}]{KohnPR1965}%
  \BibitemOpen
  \bibfield  {author} {\bibinfo {author} {\bibfnamefont {W.}~\bibnamefont
  {Kohn}}\ and\ \bibinfo {author} {\bibfnamefont {L.~J.}\ \bibnamefont
  {Sham}},\ }\bibfield  {title} {\bibinfo {title} {Self-consistent equations
  including exchange and correlation effects},\ }\href
  {https://doi.org/10.1103/PhysRev.140.A1133} {\bibfield  {journal} {\bibinfo
  {journal} {Phys. Rev.}\ }\textbf {\bibinfo {volume} {140}},\ \bibinfo {pages}
  {A1133} (\bibinfo {year} {1965})}\BibitemShut {NoStop}%
\bibitem [{\citenamefont {Bl\"ochl}(1994)}]{BlochlPRB1994}%
  \BibitemOpen
  \bibfield  {author} {\bibinfo {author} {\bibfnamefont {P.~E.}\ \bibnamefont
  {Bl\"ochl}},\ }\bibfield  {title} {\bibinfo {title} {Projector augmented-wave
  method},\ }\href {https://doi.org/10.1103/PhysRevB.50.17953} {\bibfield
  {journal} {\bibinfo  {journal} {Phys. Rev. B}\ }\textbf {\bibinfo {volume}
  {50}},\ \bibinfo {pages} {17953} (\bibinfo {year} {1994})}\BibitemShut
  {NoStop}%
\bibitem [{\citenamefont {Kresse}\ and\ \citenamefont
  {Furthm\"uller}(1996)}]{KressePRB1996}%
  \BibitemOpen
  \bibfield  {author} {\bibinfo {author} {\bibfnamefont {G.}~\bibnamefont
  {Kresse}}\ and\ \bibinfo {author} {\bibfnamefont {J.}~\bibnamefont
  {Furthm\"uller}},\ }\bibfield  {title} {\bibinfo {title} {Efficient iterative
  schemes for \textit{ab initio} total-energy calculations using a plane-wave
  basis set},\ }\href {https://doi.org/10.1103/PhysRevB.54.11169} {\bibfield
  {journal} {\bibinfo  {journal} {Phys. Rev. B}\ }\textbf {\bibinfo {volume}
  {54}},\ \bibinfo {pages} {11169} (\bibinfo {year} {1996})}\BibitemShut
  {NoStop}%
\bibitem [{\citenamefont {Kresse}\ and\ \citenamefont
  {Hafner}(1994)}]{kresse1994ab}%
  \BibitemOpen
  \bibfield  {author} {\bibinfo {author} {\bibfnamefont {G.}~\bibnamefont
  {Kresse}}\ and\ \bibinfo {author} {\bibfnamefont {J.}~\bibnamefont
  {Hafner}},\ }\bibfield  {title} {\bibinfo {title} {Ab initio
  molecular-dynamics simulation of the liquid-metal--amorphous-semiconductor
  transition in germanium},\ }\href {https://doi.org/10.1103/PhysRevB.49.14251}
  {\bibfield  {journal} {\bibinfo  {journal} {Phys. Rev. B}\ }\textbf {\bibinfo
  {volume} {49}},\ \bibinfo {pages} {14251} (\bibinfo {year}
  {1994})}\BibitemShut {NoStop}%
\bibitem [{\citenamefont {Furness}\ \emph {et~al.}(2020)\citenamefont
  {Furness}, \citenamefont {Kaplan}, \citenamefont {Ning}, \citenamefont
  {Perdew},\ and\ \citenamefont {Sun}}]{furness2020accurate}%
  \BibitemOpen
  \bibfield  {author} {\bibinfo {author} {\bibfnamefont {J.~W.}\ \bibnamefont
  {Furness}}, \bibinfo {author} {\bibfnamefont {A.~D.}\ \bibnamefont {Kaplan}},
  \bibinfo {author} {\bibfnamefont {J.}~\bibnamefont {Ning}}, \bibinfo {author}
  {\bibfnamefont {J.~P.}\ \bibnamefont {Perdew}},\ and\ \bibinfo {author}
  {\bibfnamefont {J.}~\bibnamefont {Sun}},\ }\bibfield  {title} {\bibinfo
  {title} {Accurate and numerically efficient {r2SCAN} meta-generalized
  gradient approximation},\ }\href
  {https://doi.org/10.1021/acs.jpclett.0c02405} {\bibfield  {journal} {\bibinfo
   {journal} {J. Phys. Chem. Lett.}\ }\textbf {\bibinfo {volume} {11}},\
  \bibinfo {pages} {8208} (\bibinfo {year} {2020})}\BibitemShut {NoStop}%
\end{thebibliography}

\end{document}